\documentclass[a4paper,11pt]{article}
\pdfoutput=1 

\usepackage{jheppub} 

\usepackage[T1]{fontenc} 
\usepackage[utf8]{inputenc} 
\usepackage[english]{babel} 
\usepackage{physics}
\usepackage{slashed}
\usepackage{feynmp-auto}
\usepackage{textcomp}
\usepackage[np]{numprint}
\usepackage{multirow}
\usepackage{array}
\usepackage{xcolor,soul}
\usepackage{booktabs}

\npthousandsep{\,}
\npdecimalsign{.}
\npproductsign{\times}
\npfourdigitnosep

\newcommand{\kallen}{\lambda^{\frac{1}{2}}\qty}
\newcommand{\cj}{\overline}
\newcommand{\sh}{\slashed}
\newcommand{\pd}{\partial}

\newcommand{\ra}{\rightarrow}
\newcommand{\eg}{\emph{e.g.}\ }

\newcommand{\tapi}{\textsuperscript}
\newcommand{\tped}{\textsubscript}
\mathchardef\mhyphen="2D
\definecolor{charm}{RGB}{229,30,16}
\definecolor{ps191}{RGB}{0,158,115}
\definecolor{peak}{RGB}{230,159,0}
\definecolor{nutev}{RGB}{120,120,120}
\definecolor{bluet}{RGB}{86,180,233}

\newcommand{\refeq}[1]{Eq.~(\ref{#1})}
\newcommand{\refeqs}[2]{Eqs.~(\ref{#1})~and~(\ref{#2})}
\newcommand{\refeqss}[3]{Eqs.~(\ref{#1}), (\ref{#2})~and~(\ref{#3})}
\newcommand{\reffig}[1]{figure~\ref{#1}}

\newcommand{\refsec}[1]{section~\ref{#1}}
\newcommand{\refapp}[1]{appendix~\ref{#1}}
\newcommand{\reftab}[1]{table~\ref{#1}}
\newcommand{\refref}[1]{ref.~\cite{#1}}
\newcommand{\refrefs}[1]{refs.~\cite{#1}}

\newcounter{CommentCount}
\definecolor{PB}{rgb}{0.9,0,0}
\definecolor{TB}{rgb}{0,0.7,0}

\title{Heavy Neutral Leptons from low-scale seesaws at the DUNE Near Detector}
\author[a]{Peter Ballett,}
\author[a,b]{Tommaso Boschi}
\author[a]{and Silvia Pascoli}
\affiliation[a]{Institute for Particle Physics Phenomenology, Department of Physics, Durham University,
	South Road, Durham DH1 3LE, United Kingdom.}
\affiliation[b]{Particle Physics Research Centre, School of Physics and Astronomy, Queen Mary University of London,
	Mile End Road, London E1 4NS, United Kingdom.}

\emailAdd{peter.ballett@durham.ac.uk} 
\emailAdd{tommaso.boschi@qmul.ac.uk}
\emailAdd{silvia.pascoli@durham.ac.uk}

\abstract{Heavy nearly-sterile neutrinos are a common ingredient in extensions of the
	Standard Model which aim to explain neutrino masses, like for instance in Type I
	seesaw models, or one of its variants. If the scale of the new Heavy Neutral Leptons (HNLs) is sufficiently low, 
	observable signatures can arise in a range of current and upcoming experiments, from the LHC to neutrino experiments.
	In this article, we discuss the phenomenology of sterile neutrinos in the MeV to GeV mass range, focusing on their decays.
	We embed our discussion in a realistic mass model and consider the resulting implications.
	We focus in particular on the impact on the signal of the strong polarisation
	effects in the beam for Majorana and \mbox{(pseudo-)Dirac} states, providing formulae 
	to incorporate these in both production and decay.
	We study how the Near Detector of the upcoming Deep Underground Neutrino
	Experiment can constrain HNL states by searching for their decay products inside the detector.
	We conduct a Monte Carlo background analysis for the most promising
	signatures, incorporating the detector's particle identification capabilities,
	and estimate the experimental sensitivity of DUNE to these particles.  
	We also present an estimate of the \mbox{$\nu_\tau$-derived} HNL flux at DUNE, 
	currently missing in the literature, which allows us to discuss searches for 
	HNLs at higher masses.  
} 


\preprint{IPPP/18/76}

\begin{document}
\maketitle

\flushbottom

\section{Introduction}
\label{sec:introduction}

The evidence for three neutrino flavour oscillation is well established~\cite{Fukuda:1998mi,Aharmim:2005gt} %
and can be accounted for only if neutrino mass splittings are non zero~\cite{nufit}.
This implies that neutrinos are massive and mix, forcing to consider extensions %
of the Standard Model (SM) to explain their origin. 
A simple means of doing so is to introduce the right-handed counterpart of SM neutrinos, which are %
singlet with respect to all SM gauge symmetries.
The Lagrangian includes a Yukawa coupling between these sterile states, the Higgs boson and the leptonic doublet, %
which generates Dirac mass-terms below the scale of Electroweak Symmetry Breaking (EWSB), 
and Majorana mass terms for the new singlet states.
On diagonalisation of the resulting neutrino mass matrix, the heavy neutrino states, %
commonly known as nearly-sterile neutrinos or Heavy Neutral Leptons (HNLs) in experimental contexts, %
remain mainly in the sterile neutrino direction and have sub-weak interactions suppressed by %
elements of the extended Pontecorvo-Maki-Nakagawa-Sakata (PMNS) mixing matrix. 

These states have been connected to a vast range of phenomenological behaviours and %
even to cosmological implications (for a recent review on sterile neutrinos, see e.g.\ \refref{Abazajian:2012ys}).
For instance, nearly-sterile neutrinos in the keV region are viable warm Dark Matter candidates (see e.g.\ \refref{Asaka:2005an}), %
whereas heavier HNLs could play a role in leptogenesis~\cite{Fukugita:1986hr, Covi:1996wh, Pilaftsis:1997jf, Buchmuller:1997yu, %
Pilaftsis:2003gt, Davidson:2008bu, Akhmedov:1998qx, Asaka:2005pn, Hernandez:2015wna, Hernandez:2016kel, %
Hambye:2016sby, Hambye:2017elz, Drewes:2017zyw}.
So far, some possible hints in favour of heavy neutrinos have emerged in neutrino appearance oscillation experiments, %
specifically LSND~\cite{Aguilar:2001ty} and MiniBooNE~\cite{Aguilar-Arevalo:2012fmn, Aguilar-Arevalo:2013pmq, Aguilar-Arevalo:2018gpe} %
but are disfavoured by disappearance experiments~\cite{TheIceCube:2016oqi, Adamson:2017uda, Aartsen:2017bap}, %
unless non-standard effects are present~\cite{Liao:2016reh, Liao:2018mbg, Esmaili:2018qzu, Denton:2018dqq}. Further hints 
in the same mass range have been reported for mixing with electron neutrinos in the so-called reactor anomaly~\cite{Mueller:2011nm, Mention:2011rk, Huber:2011wv, Ko:2016owz, Alekseev:2018efk} and in the less statistically significant gallium one~\cite{Abdurashitov:2005tb, Laveder:2007zz, Giunti:2006bj}.
Explanations of the MiniBooNE low energy excess invoking GeV-scale HNLs with non-standard interactions~\cite{Gninenko:2009ks, Gninenko:2010pr, %
	Masip:2012ke, Bertuzzo:2018itn, Ballett:2018ynz}, %
have also been put forward. In these models, heavy neutral fermions are produced by neutrino up-scattering in the detector and subsequently decaying into photons or electrons, which mimic an electron neutrino interaction.
The interpretation of the current experimental results is still largely debated in the scientific community.
Searches both for electron-like signatures in MicroBooNE, the SBN programme at Fermilab~\cite{Antonello:2015lea}, %
and in short baseline reactor neutrino experiments, such as DANNS~\cite{Alekseev:2018efk}, NEOS~\cite{Ko:2016owz}, %
PROSPECT~\cite{Ashenfelter:2018iov}, STEREO~\cite{Almazan:2018wln}, NEUTRINO-4~\cite{Serebrov:2018vdw}, %
will shed further light on these possibilities, %
whereas experiments like KATRIN~\cite{Mertens:2018vuu} will be able to exclude the gallium anomalies.

Apart from these controversial hints, no positive evidence of heavy neutrinos has been found to date in laboratory searches.
A thorough review of the current constraints can be found in~\refrefs{Atre:2009rg, Drewes:2015iva}. 
%
%
%
%
Bounds critically depend on the HNL masses and the flavour with which they mix.
Searches for kinks in Curie plots of $\beta$-decay
spectra~\cite{Galeazzi:2001py, Hiddemann:1995ce, Holzschuh:1999vy,
	Holzschuh:2000nj, Deutsch:1990ut} have placed bounds on the electronic
mixing for HNL masses between the keV and MeV scales.  
For masses from a few MeV to a few hundreds MeV, searches for monochromatic peaks in the lepton spectrum of decaying pions %
and kaons place important bounds on the muonic and electronic mixing angles~\cite{Artamonov:2014urb, Britton:1992pg, Britton:1992xv, Aguilar-Arevalo:2017vlf, Aguilar-Arevalo:2019owf}.
Neutrinoless double beta decay indirectly constrains Majorana HNLs from the eV to the TeV scale %
and lepton number violating meson and tau decays can be used to set limits on the mixing angle %
in narrow ranges of HNLs masses~\cite{Atre:2009rg}. 
The tightest constraints come from searches for the direct production and subsequent decays of heavy neutrinos %
in \emph{beam dump} experiments and at colliders.
The strongest limits were set by the PS191 experiment~\cite{Bernardi:1985ny, Bernardi:1987ek}, %
a beam dump experiment which ran at CERN in 1984.
Its most stringent upper bounds on the novel mixing angles are \mbox{$|U_{e4}|^2, |U_{\mu4}|^2 \leq \np{e-8} \text{\,--\,} \np{e-9}$} %
for neutrino masses between the pion and the kaon mass.
Other bounds of this type can be found in~\refrefs{Abreu:1996pa, Adriani:1992pq, Vilain:1994vg, Vaitaitis:1999wq, Badier:1985wg, CooperSarkar:1985nh, Gallas:1994xp}
as well as collider ones, from LHCb~\cite{Aaij:2014aba}, ATLAS~\cite{Aaboud:2018spl}, CMS~\cite{Sirunyan:2018mtv, Sirunyan:2018xiv}, BELLE~\cite{Liventsev:2013zz} %
(see also \refref{Harrison:2015bja}).

%

It is exciting to note that current and upcoming neutrino oscillation
experiments will be able to perform beam dump style measurements \cite{Kusenko:2004qc, Asaka:2012bb, Abe:2019kgx}. 
A crucial difference between oscillation detectors and dedicated beam dump searches of the past is that %
the former tries to maximise its Standard Model neutrino scattering rate, while the latter goes to lengths %
to suppress it in order to reduce backgrounds.
However, for some of the current and future accelerator neutrino experiments, %
such as the Short Baseline Neutrino (SBN) program~\cite{Antonello:2015lea}, %
the strong particle reconstruction capabilities of Liquid Argon detectors and distinctive kinematics %
of neutrino decays have been shown to allow competitive bounds on heavy neutrinos %
to be made despite naively large backgrounds \cite{Ballett:2016opr}. 
Long baseline oscillation experiments, such as the upcoming Deep Underground Neutrino Experiment (DUNE)~\cite{Abi:2018dnh}, %
will see a greatly diluted flux of nearly-sterile neutrinos at their far detectors and consequently poor sensitivity.
However, the DUNE Near Detector (DUNE ND), placed \np{574}\,m from the target, has a great potential %
for searches for new physics~\cite{Adams:2013qkq}.
Even if the final design of the ND has not been confirmed as yet, the options being considered combine a large active volume, %
in close proximity to a very intense neutrino beam and cutting-edge event reconstruction capabilities.
These will allow DUNE ND to undertake valuable searches for BSM physics in a entirely complementary way %
to the central oscillation physics programme.

In this article, we present a detailed analysis of the sensitivity of DUNE ND to HNL in beam dump style searches.
We ground our discussion in theoretically consistent models, in which sterile neutrinos %
are associated with neutrino mass generation via a low-scale seesaw mechanism.
We extend and refine previous analyses~\cite{Krasnov:2019kdc, Adams:2013qkq}, %
using the latest configuration of the DUNE ND~\cite{DUNETDR:2019, DUNEND:2019}.
We note that the range of masses and mixing angles testable at DUNE~ND is of interest for the generation of %
the baryon asymmetry in the context of the ASR mechanism~\cite{Akhmedov:1998qx, Asaka:2005pn, Hernandez:2015wna, Hernandez:2016kel, Drewes:2017zyw}.
We consider both Majorana and pseudo-Dirac states and calculate their decay and production rates, 
with careful consideration given to helicity arguments.
These formulae are then used to estimate the sensitivity of the experiment, %
taking into account the beam and detector performance capabilities thanks to simulations of both event and background signals.
We stress that DUNE will be able to extend the current limits on new fermionic singlets, %
including those with masses above 500\,MeV, probing models of theoretical significance for the generation of neutrino mass.
We show that bounds can be put also on the mixing with tau neutrinos, thanks to the high energy~beam.
%

The article is organised as follows.
In \refsec{sec:model} we discuss neutrino mass generation and its consequences for heavy neutrinos in the mass range of interest.
In \refsec{sec:decay} and \refsec{sec:production}, we present the nearly-sterile neutrino decay and production rates accounting for both %
Majorana/(pseudo-)Dirac states and fully incorporating helicity effects and distributional information about the final-state observables. 
In \refsec{sec:experiment}, we turn to DUNE~ND, describing our assumptions about the experimental apparatus, %
our neutrino flux modelling, including a $\nu_\tau/\cj{\nu}_\tau$ simulation, %
the expected signal and the impact of backgrounds.
In \refsec{sec:results}, we quantify the sensitivity of DUNE ND to decays of heavy neutrino and, %
in \refsec{sec:combined}, its ability to constrain the parameter space of low-scale seesaw models.
Our concluding remarks are made in \refsec{sec:conclusions}.

\section{Heavy neutrinos in seesaw models}
\label{sec:model}

The lightness of the observed neutrino masses can be explained in a range of different scenarios. %
New SM-gauge singlet fermions are a feature common to many of them. 
The~most general Lagrangian including a set of right-chiral gauge singlets 
$\{{\cal{N}}_i\}$ is given by 
\begin{equation}
	\label{eq:model}
	\mathcal{L}_{\text{SM}+\mathcal{N}} = \mathcal{L}_\text{SM} + i\, \cj{\mathcal{N}}_i \sh{\pd} \mathcal{N}_i + Y_{\alpha i} \cj{L}_\alpha \widetilde{H} \mathcal{N}_i + \frac{1}{2} (M_R)_{ij} \cj{\mathcal{N}^c}_i \mathcal{N}_j + \text{h.c.}\ ,
\end{equation}
with $\mathcal{L}_\text{SM}$ denoting the SM Lagrangian and the other symbols taking their conventional meaning.
Without loss of generality, $M_R$ can be chosen to be diagonal.
After electroweak symmetry breaking, a Dirac mass emerges for which we will use the notation $m_D \equiv v Y/\sqrt{2}$.
This term appears, for instance, in the famous Type I seesaw mechanism~\cite{Minkowski:1977sc,Mohapatra:1979ia,GellMann:1980vs,Yanagida:1979as}. %
Majorana masses for the light neutrinos arise and are given by
\begin{equation}
	\label{eq:typeI}
	m_\nu = -m_D M_R^{-1} m_D^T + \mathcal{O}\qty(\qty[m_D M_R^{-1}]^2)\ .
\end{equation}
The heavy neutrino masses are approximately given by the diagonal entries of $M_R$ and its corresponding eigenstates, %
the heavy nearly-sterile neutrinos $N_i$, have suppressed mixing with active neutrinos and are mainly composed %
by sterile fields.
Neglecting the matrix nature of these expressions for now, if $m_D$ takes values around the electroweak scale, %
acceptable neutrino masses are produced when $M_R$ has values around the GUT scale, %
suggestively connecting it to a high-scale breaking of $U(1)_{B-L}$~\cite{Minkowski:1977sc}.
Low-scale solutions are also possible by taking the Yukawa couplings to be similar %
or smaller than the other SM lepton Yukawa couplings, \eg if $m_D$ takes values in the keV range, %
new nearly-sterile states would exist with masses around a GeV.
 The resulting mixing is constrained by the contribution given to light neutrino masses and naively one can expect to have
\begin{equation}
	\label{eq:TypeI_mixing_bound}
	\qty|U_{\alpha N}|^2 \lesssim \frac{m_\nu}{m_N} \lesssim  \np{e-10}\ \frac{1\,\mathrm{GeV}}{m_N}\ ,
\end{equation}
where we have taken $m_\nu\lesssim 0.1$\,eV.
Specific models in which low energy neutrino masses and mixing angles are derived from the see-saw parameters %
allow for a broader range of values, invoking specific structures for the Yukawa couplings.
For example, it has been pointed out that in some cases a lower bound on different combinations of %
$|U_{ \alpha N}|^2$ can be found, depending on the number and mass hierarchy of heavy neutrinos %
and the value of the lightest neutrino masses (see \refrefs{Gorbunov:2013dta,Drewes:2019mhg}).
Values larger than the ones naively expected from \refeq{eq:TypeI_mixing_bound} can be obtained %
by advocating specific cancellations in the heavy neutrino contribution to light neutrino masses.
For instance, in analogy to \refref{Moffat:2018wke}, one can use the Casas-Ibarra parametrisation~\cite{Casas:2001sr}
\begin{equation}
	\label{eq:CasasIbarra}
	m_D \simeq i \ \mathcal{U} \ m_\text{diag}^{1/2}\ R\ M_R^{-1/2}\ ,
\end{equation}
where $\mathcal{U}$ is the usual $3\times 3$ PMNS mixing matrix %
which diagonalises $m_\nu$ into the diagonal matrix containing the three light neutrinos $m_\text{diag}$, %
and $R$ is an arbitrary complex orthogonal matrix.
We notice that the values of the entries of $R$ are not bounded and could be much larger than one.
Consequently, the mixing angles between the active and heavy sterile neutrinos, %
which scale as $m_D\, M_{R}^{-1}$, can be enhanced without violating the bound from neutrino masses, %
in which $R$ does not enter.
Differently from \refref{Moffat:2018wke}, one loop corrections are negligible at the GeV scale and do not introduce additional fine-tuning. 

Without advocating specific forms of the Yukawa coupling, %
the bound in Eq.~(\ref{eq:TypeI_mixing_bound}) can be avoided in presence of more than three sterile neutrinos %
thanks to the cancellation between their contributions to neutrino masses.
In recent years, a lot of interest has been focused on these type of models %
e.g.\ the Inverse Seesaw~(ISS)~\cite{Mohapatra:1986bd, GonzalezGarcia:1988rw}, extended seesaw~\cite{Barr:2003nn} %
and linear seesaw models~\cite{Malinsky:2005bi,Kang:2006sn}.
For~definiteness, we will focus on the ISS model.
In this case, a quasi-preserved lepton number guarantees the specific texture of $M_R$ and $m_D$ %
and its small breaking is natural in the 't Hooft sense~\cite{tHooft:1980xss}.

The physical spectrum of heavy neutrinos can be best understood in the Lepton Number Conserving (LNC) limit.
We use ISS\,$(a,b)$ to denote the model with $a$ $(b)$, $a,b\neq 0$, new gauge singlets of lepton number $+1$ $(-1)$.
Following \refeq{eq:model} and \refeq{eq:typeI}, the most general mass matrices are then given by 
\[
	m_D = \qty( \begin{matrix} m, 0\end{matrix}) \qquad \text{and}\qquad M_R = %
	\qty(\begin{matrix} 0 & M^\text{T} \\ M & 0 \end{matrix})\ ,
\]
where we introduce the $3\times a$ complex matrix $m$ and the $b\times a$ complex matrix $M$.
The spectrum of physical states in the LNC limit for ISS\,$(a,b)$ is given by
\[	
	\text{min}\{3+b,a\}~\text{Dirac pairs}\qquad\text{and}\qquad
	|3 + b-a|~\text{massless Weyl states}.  
\] 
The masses of the Dirac pairs are the non-zero singular values of the rectangular $a \times (3+b)$ matrix $(m^\text{T}, M^\text{T})$.
Note that for $a\neq b$, in addition to a set of Dirac pairs of arbitrary masses, extra massless sterile states %
degenerate with the light neutrinos are present.
Mixing involving these degenerate states is not defined in the LNC limit, as any unitary map in the degenerate subspace is permissible.
On the contrary, the introduction of a small Lepton Number Violating (LNV) parameter %
perturbs the LNC spectrum as well as the mixing.
In general there are only two possible origins for a low-scale heavy neutrino: 
 \begin{itemize}
		\item A massless Weyl fermion in the LNC limit which is given non-zero mass proportional to the perturbation.
			As the mixing between massless states is not defined in the LNC limit, the perturbation controls %
			the induced mixing between the nearly-sterile state and the active ones.
			We will refer to this state as a Majorana neutrino.
		\item A massive Dirac pair at the low scale in the LNC limit, becoming a pseudo-Dirac pair after the perturbation, %
			which regulates the mass splitting of the pair.
			In the LNC limit, the mixing angles between Dirac pair and light neutrinos can be arbitrarily large,
			and this property remains after the perturbation.
	\end{itemize}
	The first case only arises in models with an imbalanced number of new fields, %
	i.e.\ ISS\,$(a,b)$ such that $a\neq b$, while the second option can occur in any ISS model. 

In this paper, we are interested in heavy states with masses in the MeV--GeV range.%
\footnote{This is motivated by the kinematic limits on production from meson decays discussed in more detail in \refsec{sec:production}.} %
Our discussion above suggests that both Majorana states and (pseudo-)Dirac states should be considered, %
covering all possible phenomenological aspects.
In what follows, we will compute the production and decay rates for Majorana states %
and Dirac states and study their discovery potential at DUNE ND.
We disregard lepton number violating effects and therefore the distinction between pseudo-Dirac and Dirac states will not be relevant.

Feynman rules for Majorana states derived from \refeq{eq:model} can be found in~\cite{Atre:2009rg},
or constructed using the techniques of \refref{Denner:1992me}. 
For an explicit comparison between Dirac versus Majorana Feynman rules for heavy neutrinos, see~\refref{Abada:2016plb}.

\section{Heavy neutrino decay}
\label{sec:decay}

In this section we compute the heavy neutrino decay rates and polarised %
distributions necessary for the simulation of beam dump searches.
We compute rates for both Majorana and (pseudo-)Dirac states, allowing us to consistently %
explore the parameter space of low-scale seesaw models.
%

This analysis can be simplified by noting the following equivalences.
A Majorana neutrino $N$ decaying via a charged current process has the same differential %
decay rate as the Dirac neutrino $N_D$ with the appropriate lepton number, 
\[
	\dd{\Gamma}(N\to \ell_\alpha^- X^+) = \dd{\Gamma}(N_D\to \ell_\alpha^- X^+) \quad\text{and}\quad
	\dd{\Gamma}(N\to \ell_\alpha^+ X^-) = \dd{\Gamma}(\cj{N}_D\to \ell_\alpha^+ X^-)\ ,
\]
where we assume identical mass and mixing angles for both Dirac and Majorana neutrinos. 
This equivalence can be seen directly from the Feynman rules for Dirac and Majorana fermions~\cite{Denner:1992me} %
(see also \refref{Abada:2016plb}), but also explicitly in our formulae below.
In a neutral current (NC) decay, however, the two contractions of the NC operator lead to another contribution,  
\begin{equation*}
	\dd{\Gamma}(N \to \nu X') = \dd{\Gamma}(N_D\to \nu X') + \dd{\Gamma}(\cj{N}_D\to\cj{\nu} X')\ .
\end{equation*}
These relations hold at the differential level if the kinematic variables are reinterpreted in the obvious way.
In this sense, we can view the Majorana process as the sum of Dirac particle and antiparticle decays.%
\footnote{In a general amplitude with Majorana states, there would also be an interference contribution between these two %
	sub-processes. However, in all cases of interest, interference diagrams are proportional to the %
	final-state light neutrino mass, which we take to be zero.} 
%
%
Considering the total decay rates only, we find that the Majorana decay is larger by a factor of 2 compared to the Dirac case,
\[
	\Gamma(N\to \nu X' ) = 2\Gamma(N_D \to \nu X')\ .
\]
Note that this is true only for the total decay rates with massless final-state neutrinos.

\begin{table}
	\small
	\centering
	\begin{tabular}{lrlrlr}
		\toprule                                    
		Channel			& Threshold		    &      Channel			& Threshold		&   Channel			& Threshold		\\
		\cmidrule(lr){1-2} \cmidrule(lr){3-4} \cmidrule(lr){5-6}                                   
		$\nu \nu \nu$		& \np{e-9}\,MeV	&  $e^\mp K^\pm$	    & \np{494}\,MeV		&   $\nu \eta'$		        & \np{958}\,MeV		\\
		$\nu e^+ e^-$		& \np{1.02}\,MeV&  $\nu \eta$	        & \np{548}\,MeV		&   $\mu^\mp K^{*\pm}$		& \np{997}\,MeV		\\
		$\nu e^\pm \mu^\mp$	& \np{105}\,MeV	&  $\mu^\mp K^\pm$      & \np{559}\,MeV		&   $\nu \phi$		        & \np{1019}\,MeV		\\
		$\nu \pi^0$ 		& \np{135}\,MeV	&  $\nu \rho^0$	        & \np{776}\,MeV		&   $\nu e^\pm \tau^\mp$	& \np{1776}\,MeV		\\
		$e^\mp \pi^\pm$		& \np{140}\,MeV	&  $e^\mp \rho^\pm$	    & \np{776}\,MeV		&   $e^\mp D^\pm$           & \np{1870}\,MeV		\\
		$\nu \mu^+ \mu^-$	& \np{210}\,MeV	&  $\nu \omega$	        & \np{783}\,MeV		&   $\nu \mu^\pm \tau^\mp$	& \np{1880}\,MeV		\\
		$\mu^\mp \pi^\pm$	& \np{245}\,MeV	&  $\mu^\mp \rho^\pm$	& \np{882}\,MeV		&   $\tau^\mp \pi^\pm$		& \np{1870}\,MeV		\\
		&               &  $e^\mp K^{*\pm}$	    & \np{892}\,MeV		&                           &                       \\
		\bottomrule
	\end{tabular}
	\footnotesize
	\caption{All the available channels for a HNL with a mass below the $D_s^\pm$ mass are listed above, %
		sorted by threshold mass.
		The active neutrino is considered massless, when compared to the masses of the other particles.}
	\label{tab:decays}
\end{table}

It is instructive to reconsider this result in the light of the \emph{practical Dirac--Majorana confusion theorem}~\cite{Kayser:1981nw, %
	Kayser:1982br}.
In \refref{Kayser:1981nw}, the decomposition into particle and antiparticle processes was performed for %
Majorana neutrino--electron scattering via neutral current, leading to a factor of two enhancement in the total rate.
However, this enhancement was shown to be absent in practice due to the polarisation of the incoming neutrino, %
which suppresses the $\Delta L = 2$ contributions by factors of the neutrino mass. 
In the present case of nearly-sterile decay, where mass effects are large and essential to the calculation, %
there is no analogous effect: Dirac and Majorana neutrinos will have distinct total decay rates regardless of their polarisation. 
%
%
Therefore, the total decay rates of heavy neutrinos into observable final states could in principle allow us to determine %
the Majorana/Dirac nature of the initial state.
This is not a trivial effect: a pure Majorana state decays with equal probability into $e^-\pi^+$ as $e^+ \pi^-$, %
one of its dominant and most experimentally distinctive branching decay modes, while a Dirac heavy neutrino %
will only decay into $e^-\pi^+$.
Assuming charge-identification is possible in the detector, distinguishing between the two total decay rates %
should be possible with modest statistics. %
In a charge-blind search or for an NC channel, the total decay rate of Majorana neutrinos would appear to be twice as large as that of Dirac neutrinos.
However, being the mixing usually an unknown quantity, the difference between Majorana and Dirac nature cannot be deduced as easily.

There is also a more subtle impact of the nature of the decaying neutrino.
Even though the total decay rate is not affected by the helicity of the initial neutrino, %
the helicity does affect the distributions of final state particles, which will in turn influence the observability of %
the signatures of neutrino decay. 
It is important that these polarisation effects are correctly implemented when studying the distributions of %
final state observables and subsequently when developing an analysis to tackle backgrounds. 

\begin{figure}
	\centering
	\resizebox{\textwidth}{!}{\input{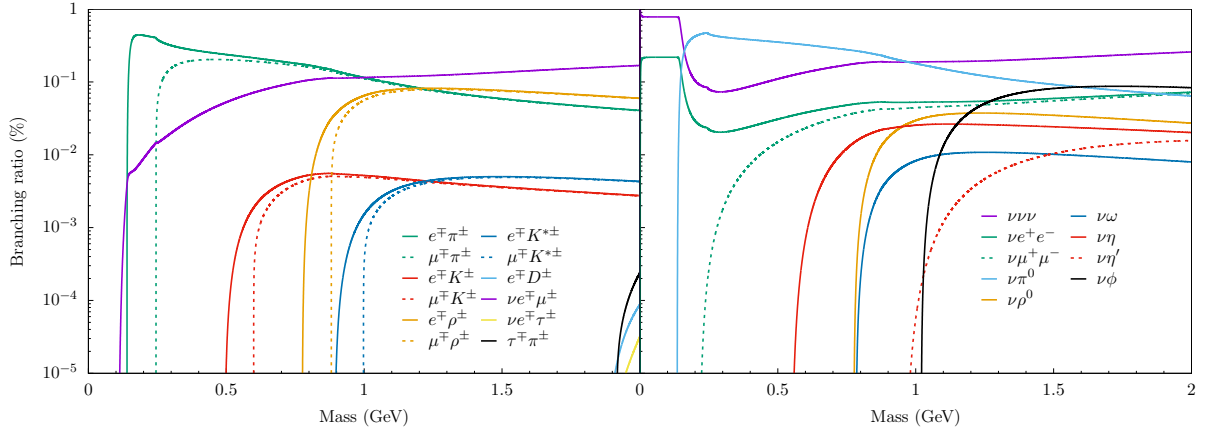}}
	\footnotesize
	\caption{The branching ratios for HNL decays, integrated over the angular variables, are shown above %
		as functions of the mass.
		They are grouped in CC-mediated decays (left) and NC-mediated decays (right), in the range from \np{0.01}\,MeV up to %
		the maximum mass limit for neutrino production, near 2\,GeV. 
		A scenario in which $|U_{e N}|^2=|U_{\mu N}|^2=|U_{\tau N}|^2$ is chosen here %
		for illustrative purposes.
		The branching ratios of Majorana neutrinos and Dirac neutrinos are mathematically identical and %
		therefore no distinction is stressed.
		The decay into three light neutrinos is fundamental for a correct computation %
		of the branching ratios, even though fully invisible from an experimental point of~view.}
	\label{fig:branch}
\end{figure}

In the remainder of this section, we present results for the polarised heavy neutrino decay rates %
and distributions for Majorana and (pseudo-)Dirac neutrinos.
The decay modes considered are listed in~\reftab{tab:decays} and the respective branching ratios as functions of %
the neutrino mass are shown in~\reffig{fig:branch}.
The differential widths have been computed using the massive spinor helicity formalism %
(see e.g.\ \refrefs{Dittmaier:1998nn, Diaz-Cruz:2016ahc}), and checked numerically using %
FeynCalc~\cite{Shtabovenko:2016sxi,Mertig:1990an}.

\subsection{Polarised Majorana neutrino decay}

Although spin-averaged Majorana neutrino decay rates are well known in the literature~\cite{Atre:2009rg,Gorbunov:2007ak,Helo:2010cw} %
(see also Ref.~\cite{Bondarenko:2018ptm}), to the best of our knowledge the polarised rates are not.
These are necessary to correctly describe the distributions of observables in a beam dump experiment, %
and in this section we present formulae for these differential decay rates. 

\newpage
Note that we stay agnostic as to the final nature and flavours of outgoing %
neutrinos, and in all cases sum over any possible outgoing states to define a %
semi-inclusive decay rate into the visible particle(s) $X'$, i.e.\ 
\[
	\Gamma(N\to \nu X') \equiv \sum_{i=1}^3 \Gamma(N\to\nu_i X')\ .
\]
The alternative, chosen by many other authors, is to treat light neutrinos as Dirac particles, %
and construct the full decay width using arguments of CP invariance, %
in practice amounting to adding some judicious factors of two~\cite{Atre:2009rg,Bondarenko:2018ptm}.
Following this approach, our summed decay rate for $N\to\nu X'$ can be seen as 
\[
	\Gamma(N\to \nu X') \equiv \sum_{\alpha=e}^\tau \qty[\Gamma(N\to\nu_\alpha X') + %
	\Gamma(N\to\overline{\nu}_\alpha X')]\ .
\]
The two approaches are identical mathematical procedures and can both be used to compute the differential decay rates; %
however, we avoid the latter as the light neutrinos in most seesaw models are Majorana fermions, and making a distinction %
between $\nu_\alpha$ and $\overline{\nu}_\alpha$ is physically misleading.%
\footnote{The approach could be seen as a short-hand for decay rates into polarised massless neutrinos, %
	but as we are particularly concerned with polarisation effects in the beam this only adds a further complication.}
We also find that the distribution of events, the role of helicity and the heavy neutrino nature are obscured by this approach.
In contrast, by summing over all outgoing states, our formulae are insensitive to the Majorana/Dirac nature of the light neutrinos, %
and are the physically relevant rates necessary for comparison with beam dump experiments, as outgoing neutrinos are not reconstructed.

\subsubsection{Pseudoscalar mesons}
\label{sec:pseudoscalar}

The semi-leptonic meson decays are some of the most important channels identified in previous studies~\cite{Atre:2009rg, %
	Ballett:2016opr} (see also \refref{Asaka:2012bb}) %
thanks to their large branching ratios and distinctive final state particles.
Both charged and neutral pseudo-scalar mesons are viable final state particles, namely $P^\pm$ and $P^0$, %
and the decay widths are given in the Centre of Mass (CM) frame by
\begin{align}
	\dv{\Gamma_\pm}{\Omega_{\ell_\alpha}} \qty(N \to \ell_\alpha^- P^+) &= %
	|U_{\alpha N}|^2|V_{q\, \cj{q}}|^2  \frac{G_\text{F}^2f_P^2 m_N^3}{16\pi} %
	\ I^\pm_1 \qty(\xi^2_\alpha, \xi^2_P; \theta_\alpha)\ ,\label{eq:M_decay_pseudo_plus}\\
	\dv{\Gamma_\pm}{\Omega_{\ell_\alpha}} \qty(N \to \ell_\alpha^+ P^-) &= %
	|U_{\alpha N}|^2|V_{q\, \cj{q}}|^2  \frac{G_\text{F}^2f_P^2m_N^3}{16\pi} %
	\ I^\mp_1 \qty(\xi^2_\alpha, \xi^2_P; \theta_\alpha)\ ,\label{eq:M_decay_pseudo_minus}\\
	\dv{\Gamma_\pm}{\Omega_P} \qty(N \to \nu P^0)\hphantom{{}^-} &= %
	\qty(\sum_{\alpha=e}^\tau| U_{\alpha N}|^2) \frac{G_\text{F}^2f_{P^0}^2m_N^3}{16\pi} %
	\ \frac{I_1 \qty(0, \xi^2_P)}{4\pi}\ ,\label{eq:M_decay_pseudo_0}
\end{align}
where $\Gamma_h$ is the decay rate for neutrinos of helicity $h$, %
$V_{q \cj{q}}$ is the appropriate CKM matrix element for the considered meson, $f_P$ is its decay constant %
and $\xi_i = m_i/m_N$ denotes the mass of the final state particle $i$ %
as a fraction of the initial state mass. 
The solid angle elements $\Omega_{\ell_\alpha}$ and $\Omega_P$ refer respectively to the charged lepton and %
pseudo-scalar meson angle with respect to the neutrino direction.
The kinematic function $I_1(x,y)$~\cite{Atre:2009rg} and its angular generalisation accounting for helicity, $I_1^\pm(x,y;\theta)$, %
are defined in \refapp{app:integrals}.
After integrating over the angular variables, we find that %
both the pseudo-scalar meson decay rates do not depend on helicity, as expected,
%
%
\begin{align}
	\Gamma_\pm \qty(N \to \ell_\alpha^- P^+) &= \Gamma_\pm \qty(N \to \ell_\alpha^+ P^-) = %
	|V_{q\, \cj{q}}|^2 |U_{\alpha N}|^2 \frac{G_F^2f_P^2m_N^3}{16\pi}\ I_1\qty(\xi^2_\alpha, \xi^2_P)\ ,\\
	\Gamma_\pm \qty(N \to \nu P^0)\hphantom{{}^-} &= \left(\sum_{\alpha=e}^\tau|U_{\alpha N}|^2\right) %
	\frac{G_F^2f_P^2m_N^3}{16\pi}\ I_1\qty(0,\xi^2_P)\ .
\end{align}
These rates agree with those presented in \refrefs{Bondarenko:2018ptm,Gorbunov:2007ak} %
(correcting a factor of two discrepancy in the $\nu P^0$ rate of \refrefs{Atre:2009rg,Helo:2010cw}).

The decay into a neutral meson, in \refeq{eq:M_decay_pseudo_0}, is isotropic in the rest frame, while the charged-pion modes, %
\refeqs{eq:M_decay_pseudo_plus}{eq:M_decay_pseudo_minus}, inherit their angular dependence from $I^\pm(x, y; \theta_\alpha)$, %
on the lepton angle to the beam line in the heavy neutrino rest frame $\theta_\alpha$.
The isotropy of the neutral current decay $N\to\nu P^0$ is a manifestation of %
the Majorana nature of the particle, in agreement with the discussion of \refref{Balantekin:2018ukw}.
It is worth noting that, if final states are not charge-identified, a similar isotropy %
is obtained for the total rate of charged semi-leptonic decays, 
\begin{align}  
	\dv{\Gamma_\pm}{\Omega_{\ell_\alpha}} \qty(N \to \ell_\alpha P) &\equiv
	\dv{\Gamma_\pm}{\Omega_{\ell_\alpha}} \qty(N \to \ell_\alpha^+ P^-) +
	\dv{\Gamma_\pm}{\Omega_{\ell_\alpha}} \qty(N \to \ell_\alpha^- P^+) \notag \\
	&= |U_{\alpha N}|^2|V_{q\, \cj{q}}|^2  \frac{G_\text{F}^2f_P^2m_N^3}{16\pi}
	\ \frac{I_1 \qty(\xi^2_\alpha, \xi^2_P)}{2\pi}\ . 
\end{align}

The formulae above apply for all pseudo-scalar mesons which are kinematically allowed.
For instance, below the $K^0$ mass, the heavy neutrino can decay only into pions, %
but above $\eta$ and $\eta'$ are allowed in the final state.
%

\subsubsection{Vector mesons}

Although only for higher masses, HNL can also decay into vector mesons $V$, %
both via charged current, $N \to \ell^\mp V^\pm$, and neutral current, $N \to \nu V^0$.
We find the following polarised differential distributions in the heavy neutrino rest frame, 
\begin{align}
	\dv{\Gamma_\pm}{\Omega_{\ell_\alpha}} \qty(N \to \ell_\alpha^- V^+) &= %
	|U_{\alpha N}|^2 |V_{q\, \cj{q}}|^2 \frac{G_\text{F}^2f_V^2 m_N^3}{16\pi} %
	\ I^\pm_2\qty(\xi^2_\alpha, \xi^2_V; \theta_\alpha)\ ,\label{eq:M_decay_vector_plus}\\
	\dv{\Gamma_\pm}{\Omega_{\ell_\alpha}} \qty(N \to \ell_\alpha^+ V^-) &= %
	|U_{\alpha N}|^2  |V_{q\, \cj{q}}|^2\frac{G_\text{F}^2f_V^2m_N^3}{16\pi} %
	\ I^\mp_2 \qty(\xi^2_\alpha, \xi^2_V; \theta_\alpha)\ ,\label{eq:M_decay_vector_minus}\\
	\dv{\Gamma_\pm}{\Omega_V} \qty(N \to \nu V^0)\hphantom{{}^-} &= %
	\qty(\sum_{\alpha=e}^\tau| U_{\alpha N}|^2 ) %
	\frac{G_\text{F}^2f_{V}^2\kappa_V^2m_N^3}{16\pi}\ \frac{I_2 \qty(0, \xi^2_V)}{4\pi}\label{eq:M_decay_vector_0}\ ,
\end{align}
where $I_2(x,y)$ and $I_2^\pm(x,y;\theta)$ are defined in \refapp{app:integrals}.
We find the total decay widths given by
\begin{align}
	\Gamma\qty(N \to \ell_\alpha^- V^+) &=  \Gamma\qty(N \to \ell_\alpha^+ V^-) = %
	|U_{\alpha N}|^2 |V_{q\,\cj{q}}|^2 \frac{G_F^2 f_V^2m_N^3}{16\pi}\ I_2\qty(\xi^2_\alpha,\xi^2_V)\ ,\\
	\Gamma\qty(N \to \nu V^0)\hphantom{{}^-} &= \left(\sum_{\alpha = e}^\tau |U_{\alpha N}|^2\right) %
	\frac{G_F^2f_V^2\kappa_V^2  m_N^3}{16\pi}\ I_2\qty(0,\xi^2_V)\ ,
\end{align}
where the constants $\kappa_V$ are combinations of the Weinberg angle, depending on the flavour structure of $V^0$ (see below).
Our charged pseudo-vector decay rates agrees with \refrefs{Bondarenko:2018ptm, Atre:2009rg, Helo:2010cw, Gorbunov:2007ak} while %
our neutral pseudo-scalar calculation agrees with the corrected version presented in \refref{Bondarenko:2018ptm}.

As with the pseudo-scalar meson decay rates, the Majorana nature leads to an isotropic decay into a neutral vector meson.
An analogous effect holds for the charged vector meson decay if we assume that the charges of final state particles %
are not distinguished.
In this case, we find the physically relevant decay distribution in the particle rest frame to be given~by
\begin{align}
	\dv{\Gamma_\pm}{\Omega_{\ell_\alpha}} \qty(N \to \ell_\alpha V) &\equiv \dv{\Gamma_\pm}{\Omega_{\ell_\alpha}} %
	\qty(N \to \ell_\alpha^- V^+) + \dv{\Gamma_\pm}{\Omega_{\ell_\alpha}} \qty(N \to \ell_\alpha^+ V^-)\ , \notag\\
	&= |U_{\alpha N}|^2 \frac{G_\text{F}^2f_V^2}{16\pi} |V_{q\, \cj{q}}|^2\, m_N^3 %
	\ \frac{I_2 \qty(\xi^2_\alpha, \xi^2_V)}{2\pi}\ .
\end{align}
There are no vector mesons lighter than the $K^0$, and these decays become relevant only for higher masses %
for which decays into $\rho^\pm$ and $K^{*\pm}$, and for the neutral mode into $\rho^0$, $\omega$, and $\phi$ would be relevant.
For these neutral particles, the $\kappa_V$ factors read
\[
	\kappa_\rho   = 1-\sin^2\theta_W \quad,\quad
	\kappa_\omega = \frac{4}{3} \sin^2\theta_W \quad,\quad
	\kappa_\phi   = \frac{4}{3} \sin^2\theta_W -1\ .
\]

\subsubsection{Charged lepton pairs}

%
We assign the momenta to the particles in the three-body decay as follows
\[
	N(k_1) \to \nu(k_2)\, \ell_\alpha^-(k_3)\,\ell^+_\beta(k_4)\ ,
\] 
and denote $k_i^2 = m_i^2$.
The five-dimensional phase space of the final-state particles can be parameterised using two scaled invariant masses,
\[
	s_1=\frac{(k_2+k_3)^2}{m_N^2} \qquad \text{and} \qquad s_2=\frac{(k_2+k_4)^2}{m_N^2}\ ,
\] 
as well as three lab-frame angular variables, $(\theta_3, \phi_3)$, giving the direction of $\ell^-_\alpha$ and $\varphi_{43}$ %
denoting the relative azimuthal angle between $\ell^-_\alpha$ and $\ell^+_\beta$. 
Although $\cos\theta_4$ is not an independent element of our parametrisation, it is a physically relevant quantity %
and we use it to simplify the presentation of the distributions below.
It can be easily related to the fundamental variables $s_1,s_2,\theta_3,\varphi_3, \varphi_{43}$.
The differential decay rate is expressed as
\begin{equation}  
	\label{eq:threebody_dist_master}
	\dd{\Gamma_\pm} = \frac{G_F^2 m_N^5}{16 \pi^3} \qty(|A_0|^2\pm|A_1|^2) %
	\dd{s_1} \dd{s_2}\, \frac{\dd[2]{\Omega_3}}{4\pi}\ \frac{\dd{\varphi_{43}}}{2\pi}\ ,
\end{equation}
where $\Omega_3$ assumes the conventional meaning and with
\begin{align}
	|A_0|^2 &\equiv C_1 \qty(s_2-\xi^2_3) \qty(1+\xi^2_4-s_2) + C_2 \qty(s_1-\xi^2_4) \qty(1+\xi^2_3-s_1) \notag \\
	\label{eq:threebody_1}
	&\qquad + 2\,C_3\,\xi_3\,\xi_4 \qty(s_1+s_2 - \xi^2_3 - \xi^2_4)\ , \\
	|A_1|^2 &\equiv \qty[ C_4 \qty(s_2-\xi_3^2) - 2\,C_6\,\xi_3\,\xi_4]\kallen(1,s_2,\xi_4^2)\cos\theta_4 \notag \\
	\label{eq:threebody_2}
	&\qquad + \qty[C_5 \qty(s_1-\xi_4^2) - 2\,C_6\,\xi_3\,\xi_4]\kallen(1,s_1,\xi_3^2)\cos\theta_3\ .   
\end{align}
%
%
The coefficients $\{C_i\}$ are polynomials in chiral couplings and extended PMNS matrix elements, %
and are given for the decays of interest in \refapp{app:threebody_dist}. 
On integration over the angular coordinates, however, only the $|A_0|^2$ terms remain %
and we recover the standard expression for the total decay rates through the %
identities given in \refeqss{eq:threebody_int1}{eq:threebody_int2}{eq:threebody_int3}. 
The general expression for the total decay rate is again helicity independent and can be written as
\begin{equation}
	\Gamma_\pm = \frac{G_F^2 m_N^5}{192 \pi^3}\,\qty[ C_1\ I_1\qty(0,\xi_3^2,\xi_4^2) + %
	C_2\ I_1\qty(0,\xi_4^2,\xi_3^2) + C_3\ I_2\qty(0,\xi_3^2,\xi_4^2) ]\ .
\end{equation}
The functions $I_1\qty(x,y,z)$ and $I_2\qty(x,y,z)$ are given in \refapp{app:integrals}. 
Using the expressions for $\{C_i\}$ in \refapp{app:threebody_dist}, %
we find that the total decay rates are given to first order in the heavy-active mixing parameters $U_{\alpha N}$ by
\begin{align}
	&\Gamma_\pm\qty(N \to \nu \ell_\alpha^- \ell_\beta^+) = 
	\frac{G_F^2m_N^5}{192\pi^3}\ \qty[ |U_{\alpha N}|^2\ I_1 \qty(0, \xi_\alpha^2, \xi_\beta^2) + %
	|U_{\beta N}|^2\ I_1\qty(0, \xi_\beta^2, \xi_\alpha^2)]\ ,\\
	&\Gamma_\pm\qty(N \to \nu \ell_\alpha^- \ell_\alpha^+) = %
	\frac{G_F^2 m_N^5}{96\pi^3} \sum_{\gamma=e}^\tau |U_{\gamma N}|^2 %
	\Big\{\qty(g_L g_R + \delta_{\gamma \alpha} g_R)\  I_2 \qty(0, \xi_\alpha^2, \xi_\alpha^2)  \notag \\
	&\hspace{15em}+ \qty[g_L^2 + g_R^2 + \delta_{\gamma \alpha} (1+2g_L)]%
	\ I_1 \qty(0, \xi_\alpha^2, \xi_\alpha^2)\Big\}\ . 
\end{align}	
where $\alpha \neq \beta$, $g_L = -1/2 + \sin^2\theta_\text{W}$ and $g_R =\sin^2\theta_\text{W}$.
Our total decay rates agree with those in \refrefs{Bondarenko:2018ptm, Atre:2009rg, Helo:2010cw, Gorbunov:2007ak} %
and correct a typographical mistake in the rates presented in \refref{Ballett:2016opr}. 

All possible combinations of charged leptons except $\nu \tau^- \tau^+$ are allowed for masses below~$m_{D_s}$.
However, because of the limited phase space, the decays into $\nu \tau^\mp e^\pm$ and $\nu \tau^\mp \mu^\pm$ can be neglected.	

\subsubsection{Other decays}

There are some other decay rates relevant to this study but not as viable observable channels.
First, the total decay width of the process $N \to \nu \bar{\nu} \nu$, mediated by the $Z$ boson, reads
\begin{equation}
	\Gamma\qty(N \to \nu \bar{\nu} \nu) = \left(\sum_{\gamma = e}^\tau|U_{\gamma 4}|^2\right) \frac{G_F^2m_N^5}{96\pi^3}\ .
\end{equation}
Although this decay mode is experimentally invisible, it is the dominant channel up to the pion mass, %
when two-body semi-leptonic decays open up, and plays a significant role in defining the branching ratios of the observable channels.
Our expression agrees with \refrefs{Bondarenko:2018ptm,Atre:2009rg,Helo:2010cw,Gorbunov:2007ak}.
Secondly, there are other decay modes with small branching ratios and/or complicated final states which we do not study further.
These include the one-loop decay into a photon which has received some interest as an observable signature %
in non-minimal models~\cite{Gninenko:2009ks,Gninenko:2010pr,Magill:2018jla} where it may be enhanced. %
In the mass models considered in this work, it has a branching ratio smaller than $10^{-3}$ and will not be considered. 
%
%
%
We also neglect the multi-pion decay modes discussed in \refref{Bondarenko:2018ptm}, %
which are estimated to have at most a percent level branching ratio and a challenging hadronic final state for reconstruction. 

\subsection{Polarised (pseudo-)Dirac neutrino decay}

In this section we compute the decay rates for pseudo-Dirac pairs.
It is unlikely that any effect driven by the LNV parameter will be relevant for the discovery potential of DUNE ND %
and the signatures of these particles will be dominated by the leading order LNC effects.
Accordingly, we take the strict Dirac limit in our calculations, rather than treating the states as pseudo-Dirac pairs.

\subsubsection{Dirac (anti)neutrino decays}

The decay rates for a Dirac heavy (anti)neutrino are similar in form to those presented for the Majorana neutrino.
The key differences are lepton number conservation, which acts to forbid certain channels, and differences in %
the angular distributions of the neutral current decays.
For charged current--mediated processes, the distributions for Dirac neutrinos and antineutrinos %
are mathematically identical to the distributions for Majorana neutrinos.
The two-body semi-leptonic decays are the same of \refeqs{eq:M_decay_pseudo_plus}{eq:M_decay_vector_plus},
\begin{align}
	\dv{\Gamma_\pm}{\Omega_{\ell_\alpha}} \qty(N_D \to \ell_\alpha^- P^+) &= %
	\dv{\Gamma_\mp}{\Omega_{\ell_\alpha}} \qty(\cj{N}_D \to \ell_\alpha^+ P^-) = %
	\dv{\Gamma_\pm}{\Omega_{\ell_\alpha}} \qty(N \to \ell_\alpha^- P^+) \ ,   \\
	%
	\dv{\Gamma_\pm}{\Omega_{\ell_\alpha}} \qty(N_D \to \ell_\alpha^- V^+) &= %
	\dv{\Gamma_\mp}{\Omega_{\ell_\alpha}} \qty(\cj{N}_D \to \ell_\alpha^+ V^-)  = %
	\dv{\Gamma_\pm}{\Omega_{\ell_\alpha}} \qty(N \to \ell_\alpha^- V^+)\ .
	%
\end{align}
The situation for NC processes is different with respect to Majorana neutrinos.
The distribution of the final state particles is not isotropic anymore and it depends on the helicity state of the initial neutrino, %
in the way shown by the following differential rates
%
%
\begin{align}
	\dv{\Gamma_\pm}{\Omega_P} \qty(N_D \to \nu P^0) &= %
	\dv{\Gamma_\mp}{\Omega_P} \qty(\cj{N}_D \to \cj{\nu} P^0) = %
	\left(\sum_{\alpha=e}^\tau| U_{\alpha N}|^2\right)\frac{G_\text{F}^2f_{P^0}^2m_N^3}{32\pi} %
	\ I_1^\pm \qty(0, \xi_P^2; \theta_P)\ ,\\
	\dv{\Gamma_\pm}{\Omega_V} \qty(N_D \to \nu V^0) &= %
	\dv{\Gamma_\mp}{\Omega_V} \qty(\cj{N}_D \to \cj{\nu} V^0) = %
	\left(\sum_{\alpha=e}^\tau| U_{\alpha N}|^2\right)\frac{G_\text{F}^2f_{V}^2\kappa_V^2m_N^3}{32\pi} %
	\ I^\mp_2 \qty(0, \xi_V^2;\theta_V)\ .
\end{align}

For the three-body leptonic decays, the distribution is expressed in \refeq{eq:threebody_dist_master} %
with the relevant coefficients from \refapp{app:threebody_dist}.
The total decay rates are found to be
\begin{align}
	\Gamma_\pm\qty(N_D \to \nu \ell_\alpha^- \ell_\beta^+) &= |U_{\alpha N}|^2\frac{G_F^2m_N^5}{192\pi^3} %
	\ I_1 \qty(0, \xi_\alpha^2, \xi_\beta^2)\ ,\\ 
	\Gamma_\pm\qty(\cj{N}_D \to \cj{\nu} \ell_\alpha^- \ell_\beta^+) &= |U_{\beta N}|^2\frac{G_F^2m_N^5}{192\pi^3} %
	\ I_1 \qty(0, \xi_\beta^2, \xi_\alpha^2)\ ,\\ 
	\Gamma_\pm\qty(N_D \to \nu \ell_\alpha^- \ell_\alpha^+) &= %
	\frac{G_F^2 m_N^5}{192\pi^3} \sum_{\gamma=e}^\tau |U_{\gamma N}|^2 %
	\Big\{\qty(g_L g_R + \delta_{\gamma \alpha} g_R)\ I_2 \qty(0, \xi_\alpha^2, \xi_\alpha^2) \notag \\
	&\hspace{8em}+ \qty[g_L^2 + g_R^2 + \delta_{\gamma \alpha} \qty(1+2g_L)] %
	\ I_1 \qty(0, \xi_\alpha^2, \xi_\alpha^2)\Big\}\ ,\\ 
	\Gamma_\pm\qty(\cj{N}_D \to \cj{\nu} \ell_\alpha^- \ell_\alpha^+) &= %
	\Gamma_\mp\qty(N_D \to \nu \ell_\alpha^- \ell_\alpha^+)\ ,
\end{align}	
where $\alpha \neq \beta$, $g_L = -1/2 + \sin^2\theta_\text{W}$ and $g_R =\sin^2\theta_\text{W}$.
Our total decay rates agree with those in \refrefs{Bondarenko:2018ptm, Atre:2009rg, Helo:2010cw, Gorbunov:2007ak}.  

All decay rates not listed above are forbidden for a Dirac (anti)particle as the combination of production %
and decay would amount to a LNV process.
For the available modes, all NC modes are smaller by a factor of two for a Dirac (anti)neutrino compared %
to the equivalent Majorana process; however, the major difference we see between the Dirac (anti)neutrino and Majorana distributions %
is that these NC channels are dependent on the angular variables.
These differences in the distributions of the final state particles could be in principle exploited to identify %
the fermionic nature of the decaying HNL~\cite{Balantekin:2018ukw}.

\section{Heavy neutrino production}
\label{sec:production}

Heavy neutrinos can be produced in a beam dump experiment via the same processes %
that generate light neutrinos.
A proton beam hitting a fixed target typically yields a large number of pions and kaons, %
and also heavier mesons, the amount of which depends on the energy of the protons and the target choice.
A set of magnetic horns is responsible for the focusing of charged pions into a decay pipe; %
the other short-lived particles are usually unaffected by the deflection.
All these secondary particles decay leptonically or semi-leptonically via weak interactions thus creating a neutrino beam.
In the standard case of light neutrinos, pions and kaons principally decay into $\nu_\mu$ because two-body electronic modes are disfavoured %
by helicity suppression.
Muons decay in turn into equal numbers of $\nu_e$ and $\cj{\nu}_\mu$.
Other production sources of $\nu_e$ are the three-body decays of $K^0$ and $K^+$.
Above the neutral kaon mass, the first sizeable source of neutrino is given by the $D_s$ meson, %
for which helicity suppression again favours the production of heavy charged-leptons, %
and so $\tau$~leptons and $\nu_\tau$ are more likely to be emitted than the other flavours.
Each of the subsequent $\tau^+$ decays produces~a~$\cj{\nu}_\tau$.
We consider only the four most probable decay modes of the $\tau$ lepton, %
as they provide a sufficient description of their contribution to the overall flux.%
\footnote{The decay $\tau^+ \ra \cj{\nu}_\tau \pi^+ \pi^0$ is studied only at the level of %
	phase space sampling in this work.}

If kinematically allowed, heavy neutrino states can be sourced from these decays of mesons and charged leptons.
We show in~\reftab{tab:branch} all the neutrino production channels considered in this analysis, %
reporting the heaviest neutrino mass $m_N$ that is accessible by kinematics.
The neutrino mass range we consider goes from a few MeV up to the $D_s$ meson mass.
To~estimate the flux of heavy neutrinos produced, we start from the flux of light neutrinos, %
scaling it by an energy-independent kinematic factor.
Given a certain SM neutrino production process, $Q \ra \nu_\alpha Q'$, %
we use as scale factor the ratio between the decay width of the same process producing massive neutrinos, %
$Q \ra N Q'$, and the rate of the SM decay with light neutrinos.
The full flux of nearly-sterile neutrinos with a given helicity is therefore a linear combination of the different neutrino flux components, %
$\phi_{Q \ra\nu_\alpha}$, summing over all existing parents and all allowed flavours:
\begin{equation}
	\dv{\phi_N^\pm}{E} (E_N) \approx \sum_{Q, \alpha}  \mathcal{K}^{Q,\alpha}_\pm(m_N)\ \dv{\phi_{Q\ra\nu_\alpha}}{E} (E_N - m_N)\ , %
\end{equation}
where
\[
	\mathcal{K}^{Q,\alpha}_\pm(m_N) \equiv \frac{\Gamma_\pm(Q \ra N Q')}{\Gamma(Q \ra \nu_\alpha Q')}\ .
\]
The ratio $\mathcal{K}$ is proportional to the mixing parameter $|U_{\alpha N}|^2$ and contains only kinematic %
functions of the involved masses.
These are responsible for correcting phase space and helicity terms.

\begin{table}[t]
	\small
	\centering
	\begin{tabular}{ccrr}
		\toprule
		& Channel	& BR (\%)	& $m_N$(MeV) \\
		\hline
		$\pi^+\ra$	& $\mu^+ \nu_\mu$	& \np{99.98}		& \np{33.91}	\\
		& $e^+ \nu_e$		& \np{0.01}		& \np{139.06}	\\
		\hline
		$K^+\ra$	& $\mu^+ \nu_\mu$	& \np{63.56}		& \np{387.81}	\\
		& $\pi^0 e^+ \nu_e$	& \np{5.07}		& \np{358.19}	\\
		& $\pi^0 \mu^+ \nu_\mu$	& \np{3.35}		& \np{253.04}	\\
		& $e^+ \nu_e$		& \np{0.16}		& \np{493.17}	\\
		\hline
		$K^0_L\ra$	& $\pi^\pm e^\mp\nu_e$		& \np{40.55}	& \np{357.12}	\\
		& $\pi^\pm\mu^\mp\nu_\mu$	& \np{27.04}	& \np{252.38}	\\ 
		\hline
		$\mu^+\ra$	& $\cj{\nu}_\mu e^+ \nu_e$	&\np{100.00}	& \np{105.14}	\\
		\bottomrule
	\end{tabular}
	\hspace{3em}
	\begin{tabular}{ccrr}
		\toprule
		& Channel	& BR (\%)	& $m_N/\text{MeV}$\\
		\hline
		$D_s^+\ra$	& $\tau^+ \nu_\tau$	& \np{5.48}		& \np{191.42}	\\
		& $\mu^+ \nu_\mu$	& \np{0.55}		& \np{1862.63}	\\
		& $e^+ \nu_e$		& \np{0.008}		& \np{1967.78}	\\
		\hline
		$\tau^+\ra$ & $\pi^+\pi^0\cj{\nu}_\tau$ 	& \np{25.49}	& \np{1502.31}	\\
		& $\cj{\nu}_\tau e^+ \nu_e$ 	& \np{17.82}	& \np{1776.35}	\\
		& $\cj{\nu}_\tau \mu^+ \nu_\mu$	& \np{17.39}	& \np{1671.20}	\\
		& $\pi^+ \cj{\nu}_\tau$ 	& \np{10.82}	& \np{1637.29}	\\
		\bottomrule
	\end{tabular}
	\footnotesize
	\caption{Production channels at beam dump facilities yielding neutrinos, with the respective branching %
		ratios (taken from \refref{PDG}).
		The last column shows the maximum neutrino mass allowed if a massive state is produced.
		On the left, all the decays yielding $\nu_e$, $\nu_\mu$, and $\cj{\nu}_\mu$ up to the $K^0$ mass are shown.
		On the right, the neutrino sources which depends on the $D_s^+$ decay chain are shown; only the first four %
		decays of the $\tau$ lepton are considered in this work. }
	\label{tab:branch}
\end{table}

The helicity state plays a fundamental role in the production rate, in contrast with the case of neutrino decays, %
since there is no arbitrariness in the polarisation direction this time: %
it is defined by the neutrino momentum in the rest frame of the parent particle.
We employ the massive spinor helicity formalism to compute the production decay rates for both Majorana and Dirac neutrinos, %
and these are used to build the scale factors for each neutrino helicity.
Even though lepton number is preserved differently in the two cases and different Feynman rules hold, %
all the production channels of interest in this work are mediated by charge currents %
and therefore the rates are mathematically identical for Majorana and Dirac neutrinos.
If the neutrino is Dirac, the production decay width for an antineutrino with given helicity is the same %
as the one of the neutrino, but with opposite helicity.
The phenomenology of the scale factors is different for two-body decays and three-body decays and therefore we group them, %
respectively, in~\refsec{sec:production_2body} and~\refsec{sec:production_3body}.

\subsection{Two-body decays}
\label{sec:production_2body}

A massless neutrino (antineutrino) has its chiral and helicity states degenerate, and so %
it is always produced with a negative (positive) helicity.
It follows that the component of the light neutrino beam produced in two-body decays of pseudo-scalar mesons is polarised.
The initial spin, which is zero, must be preserved in the decay, and since the helicity of the neutrino in the rest frame %
is fixed, the accompanying lepton is produced with a ``wrong'' helicity.
This is permitted by the non-zero mass of the charged lepton and %
therefore final states with light flavour leptons undergo helicity suppression.
As soon as the neutrino mass deviates from zero, the correspondence between chirality and helicity is lost %
and the neutrino can be produced with both polarisations.
The main consequence is that the production of heavy neutrinos from light flavour mixings (electron) appears to be enhanced with %
respect to heavy flavours (muon and tau).
The effect is particularly dramatic when the mass difference between parent meson and charged lepton widens, %
as it happens with the electron decay of $D_s$, the enhancement of which is around \np{e6} for neutrino masses near 1\,GeV.

\begin{figure}[t]
	\centering
	{\resizebox{\linewidth}{!}{\input{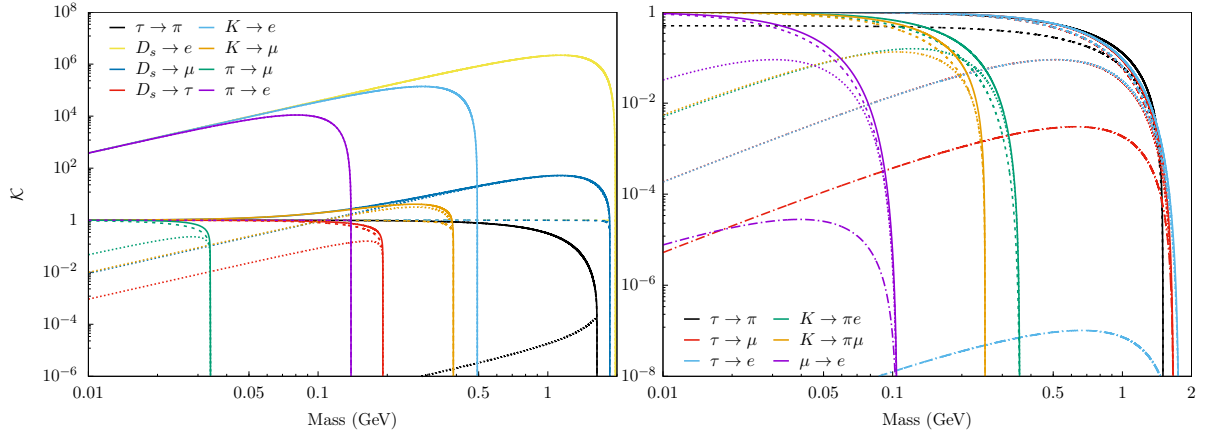}}}
	\footnotesize
	\caption{The scale factors separated by helicity components are shown.
		In two-body decays (left), the $h=-1$ components (dashed) for all channels %
		do not depend on the mass.
		The enhancement is driven by the $h=+1$ components (dotted), %
		which are the dominant contribution of the unpolarised factors (solid).
		In three-body decays (right), there are two different scale factors for purely leptonic decays, noted as $\ell_\alpha \ra \ell_\beta$: %
		if the decay is mediated by $|U_{\beta N}|^2$, for which the $h=-1$ (dashed) and the $h=+1$ (dotted) components %
		are comparable, and if the decay is mediated by $|U_{\alpha N}|^2$, %
		for which $h=-1$ dominates over the $h=+1$ (dotdashed).
		In both cases, the two parts sum up to the same quantity (solid).
		The kaon decays are also divided in $h=-1$ (dashed) and $h=+1$ (dotted) components; $\tau^+\ra\nu\pi^+\pi^0$ is %
		studied only at the phase space~level.}
	\label{fig:scale}
\end{figure}

The scale factor $\mathcal{K}_h$ for leptonic decays of a pseudo-scalar meson $P$ into neutrinos with helicity $h$, 
is given by the analytic expression:
\begin{equation}
	\label{eq:shrock_helix}
	\mathcal{K}_\pm^{P, \alpha}(m_N) = |U_{\alpha N}|^2 %
	\frac{\kallen(1,\ \xi_N^2,\ \xi_{\ell_\alpha}^2) \qty[\xi_{\ell_\alpha}^2+\xi_N^2-(\xi_N^2-\xi_{\ell_\alpha}^2)^2 %
		\pm(\xi_N^2-\xi_{\ell_\alpha}^2) \kallen(1,\ \xi_N^2,\ \xi_{\ell_\alpha}^2) ] }%
	{2 \xi_{\ell_\alpha}^2(1-\xi_{\ell_\alpha}^2)^2}\ ,
\end{equation}
where $\lambda$ is the K\"all\'en function:
\[
	\lambda(a, b, c) = (a-b-c)^2-4\,b\,c\ ,
\]
and $\xi_i = \flatfrac{m_i}{m_X}$ is the mass ratio of the final state particle $i$ over the parent particle mass.
When summing over the helicity states, the resulting factor coincides with the one computed in~\refref{Shrock:1981wq}:
\begin{equation*}
	\mathcal{K}^{P, \alpha}(m_N) = \sum_{h = \pm 1} \mathcal{K}_h^{P, \alpha}(m_N) = |U_{\alpha N}|^2 %
	\frac{\kallen(1,\ \xi_N^2,\ \xi_{\ell_\alpha}^2) \qty[\xi_{\ell_\alpha}^2+\xi_N^2-(\xi_N^2-\xi_{\ell_\alpha}^2)^2]}%
	{\xi_\alpha^2(1-\xi_{\ell_\alpha}^2)^2}\ .
\end{equation*}
In order to understand the effect of~\refeq{eq:shrock_helix}, it is convenient to define the fraction of neutrinos produced with a certain helicity as
\begin{equation*}
	S_\pm = \frac{\mathcal{K}_\pm^{P, \alpha}}{\mathcal{K}_{+1}^{P, \alpha} + \mathcal{K}_{-1}^{P, \alpha}} = 
	\frac{1}{2} \left[ 1 \pm \frac{(\xi_N^2-\xi_{\ell_\alpha}^2)\ \kallen(1,\ \xi_N^2, %
			\ \xi_{\ell_\alpha}^2)}{\xi_{\ell_\alpha}^2+\xi_N^2-(\xi_N^2-\xi_{\ell_\alpha}^2)^2}\right].
\end{equation*}
In the limit of a massless neutrino, i.e.\ $\xi_N \ra 0$, the fractions are $S_+ \ra 0$ and $S_- \ra 1$, as expected: %
all neutrinos are produced with a negative helicity.
The opposite is true when the charged lepton is in the massless limit, and the neutrinos are produced with a positive~helicity.

The only two-body decay of a lepton considered in this work is $\tau \ra \nu_\tau \pi$, and the scale factor is:
\begin{equation}
	\mathcal{K}_\pm^{\tau, \pi}(m_N) = |U_{\tau N}|^2 %
	\frac{\kallen(1,\ \xi_N^2,\ \xi_\pi^2) \qty[(1-\xi_N^2)^2 + \xi_\pi^2(1+\xi_N^2) \mp %
		(1 - \xi_N^2) \kallen(1,\ \xi_N^2,\ \xi_\pi^2) ] }%
	{2 (1-\xi_\pi^2)^2}\ .
\end{equation}
The structure is similar to the scale factor for pseudo-scalar meson two-body decays, given in~\refeq{eq:shrock_helix}, %
and analogous considerations as above can be deduced.
This is explained by crossing symmetries, as the matrix element of the process is the same.
In this case, however, the positive helicity component does not lead to any enhancement before %
the phase space~cut-off.

The effect of the scale factors as a function of the neutrino mass can be appreciated in~\reffig{fig:scale}, %
where not only helicity terms are corrected, resulting in an enhancement of the production, but also the phase space %
is properly adjusted.

\subsection{Three-body decays}
\label{sec:production_3body}

Scale factors for three-body decays are defined in the same way as two-body decay ones.
Because of the different number of degrees of freedom, the helicity of the outgoing neutrinos is not fixed by %
the spin of the parent particles.
Hence, these factors are not responsible for any enhancement in the decay rate, %
but they only quench the process as the neutrino mass upper limit is approached (see~\reftab{tab:branch}).
The scale factors have nonetheless distinct behaviours depending on the helicity state involved.
Their behaviour is plotted as a function of the heavy neutrino mass in~\reffig{fig:scale}.

The decay of a charged lepton (antilepton) of flavour $\alpha$ to a charged lepton (antilepton) of flavour $\beta$ %
can be proportional to either $|U_{\alpha N}|^2$ or $|U_{\beta N}|^2$, producing a heavy Dirac neutrino (antineutrino) in the first case %
or an antineutrino (neutrino) in the second case.
If the neutrino is Majorana, the decay can occur via both mixing matrix elements because lepton number can be violated.
Decays of muons and taus yield massive neutrinos with the following decay rate
%
%
\begin{align}
	&\Gamma_\pm (\ell_\alpha^+ \ra \ell_\beta^+ \nu N) = %
	\frac{G_F^2 m_\alpha^5}{192 \pi^3} \qty[ |U_{\alpha N}|^2\ I^\pm_\ell\qty(\xi_N^2, \xi_{\ell_\beta}^2, 0) + %
	|U_{\beta N}|^2\ I^\pm_{\cj{\ell}}\qty(0, \xi_{\ell_\beta}^2, \xi_N^2)]\ ,
\end{align}
where the integrals $I_{\ell, \cj{\ell}}(x, y, z)$ are given in~\refapp{app:integrals}.
The helicity decompositions in $I_\ell$ and $I_{\cj{\ell}}$ are different, %
but the spin-averaged decay width is the same. 

Neutral and charged kaons produce neutrinos in three-body semi-leptonic decays.
Both of them can decay into either a muon or an electron %
and a charged pion if the decaying kaon is neutral or a neutral pion if the kaon is charged.
The decay width of a pseudo-scalar meson $h_1$ to a lighter meson $h_2$ is given by
\begin{equation}
	\Gamma_\pm (h_1^{+,0} \ra h_2^{0,+}\,\ell_\alpha^+ N) = \frac{G_F^2 m_h^5}{128 \pi^3} |U_{\alpha N}|^2 |V_{q\cj{q}}|^2\ %
	I^\pm_{h_1} \qty(\xi_{h_2}^2, \xi_{\ell_\alpha}^2, \xi_N^2)\ .
\end{equation}
The integral $I_h(x, y, z)$ is reported in~\refapp{app:integrals} and consists of a combination %
of kinematic elements with terms of hadronic form factors as coefficients.
The scale factor was checked numerically against the result of~\refref{Abada:2013aba}.

The final three-body decay studied in this work is $\tau^+ \ra \cj{\nu}_\tau \pi^+ \pi^0$, %
however this channel is introduced only at the phase space level.
The scale factors for the two helicity components are therefore assumed to be identical, %
$\mathcal{K}_\pm = \smash[b]{\frac{1}{2}}$, such that the neutrino flux sub-component coming from this decay %
consists of equal number of heavy neutrinos with helicity $h=+1$ and $h=-1$.

\section{Simulation of events at DUNE ND}
\label{sec:experiment}

DUNE~\cite{Abi:2018dnh} is a long-baseline oscillation experiment that will study neutrino physics in great detail, %
focusing mainly on the determination of the CP violating phase, $\delta_\text{CP}$, %
of the mass ordering, %
and on the precision measurement of other oscillation parameters, in particular $\theta_{23}$.
These goals can be achieved thanks to both an intense neutrino beam and a high-resolution Far Detector (FD), %
consisting of a 40\,kt Liquid Argon Time Projection Chamber (LArTPC), situated \np{1300}\,km from the beam target.
The drift velocity of ionised electrons in LAr, typically of the order of cm/\textmu s, %
can be controlled with sufficient precision, by tuning the electric field, %
to result in high spatial resolution for event reconstruction~\cite{Rubbia:1977zz}.
A very sensitive FD alone, however, is not enough due to numerous uncertainties on neutrino flux and cross sections.
A smaller and closer detector, called Near Detector (ND), is therefore adopted to normalise the flux of neutrinos reaching the FD and to help cancel out many of %
the neutrino-nucleon cross section systematics.

The DUNE ND will be placed \np{574}\,m from the target.
Its definitive design has not been finalised yet, but it will likely be a hybrid concept, %
consisting of a small LArTPC placed in front of a magnetised high-pressure gaseous TPC~\cite{DUNETDR:2019, DUNEND:2019}.
This module is complementary to the front detector, controlling escaping or below-threshold particles from the LArTPC, %
but is also capable of performing standalone measurement.
For its versatile nature, it is called Multi-Purpose Detector (MPD).
The sub-system LArTPC/MPD will be movable inside the ND hall---following the DUNE-PRISM concept---for %
better profiling the neutrino flux at different angles.
There will be a third module, a 3D Scintillation Tracker (3DST), on-axis, to monitor %
the stability of the beam flux and neutron contamination.
Currently, the proposed fiducial volume for the LArTPC module is 36\,m\tapi{3} and 50\,t of LAr, %
employing the ArgonCube technology~\cite{Amsler:1993255}, %
whereas the design for the MPD is based on the TPC in \mbox{ALICE}~\cite{Glassel:2004jv}, %
a cylinder of 102\,m\tapi{3} with gas at a pressure of 10\,atm and a fiducial mass of~1\,t.
The gas assumed for the studies in the TDR is a an 80--20 mixture of Ar--CH\tped{4}.
The~3DST is designed to have a fiducial mass of around \np{8.7}\,t of plastic scintillating material and %
wavelength shifting plates.
For this analysis, we take only in account the two core sub-detectors, the LArTPC and the MPD.
The main difference between these two ND modules is that the gaseous TPC has a larger volume than the LArTPC.
This feature is favourable when studying rare events, like heavy neutrino decays, because more neutrinos enter the fiducial volume.
Furthermore, the lower density of the MPD helps reduce the number of neutrino scattering events, which are background to rare signatures.
Apart from volume and density differences and relative positions in the detector hall, %
we treat the two ND units as with similar detection performances, on-axis, and do not take in account the magnetisation of the gaseous~TPC. 

Thanks to its proximity to the accelerator, the ND will be exposed to an extremely intense neutrino beam, %
with a flux peak around five million times greater than at the FD.
The Long Baseline Neutrino Facility (LBNF) at Fermilab will deploy a very energetic beam of protons, %
extracted from the Main Injector (MI) and delivered to a graphite target.
The collision produces secondary particles, which are collimated by a focusing horn system and then decay forming a neutrino beam.
Assuming an 80\,GeV proton beam at 1.2\,MW for the first six years and at 2.4\,MW for a second set of six years~\cite{Abi:2018dnh}, 
the ND will collect a total of \np{2.65e22} protons on target (POT) over the lifespan of the experiment, %
running for the same amount of time in neutrino and antineutrino mode.
The ND will be placed on-axis for half of the total runtime, whereas it will be positioned %
at different angles off-axis for the remaining acquisition period, enacting the DUNE-PRISM concept.
The search for HNL decays can benefit to some extent at off-axis angles, %
as the SM neutrino background is particularly reduced, despite a reduced event rate.
However, the modelling of the neutrino beam profile at different angles using only the on-axis spectrum is not trivial.
Half of the total statistics will be collected with a reversed horn current configuration, %
but the parentage composition of the neutrino spectrum with the antineutrino-mode beam is not available to us, %
as well as the off-axis beam flux.
In this work, we simply consider the on-axis configuration of the ND with a forward horn current configuration, %
which would correspond to a quarter of the runtime, or \np{0.66e22}\,POT.
The same analysis of this study can nonetheless be applied equally to the beam in antineutrino mode, %
which should result in a sensitivity similar to the neutrino mode configuration, %
being wary of the different composition of the neutrino spectrum.
Even though we cannot achieve an accurate estimate of the DUNE ND sensitivity, %
we make the naive assumption that, with the above caveats, the total sensitivity to HNL%
---including off-axis angles and antineutrino mode beam---%
is equivalent to six years of data taking, i.e.\ \np{1.32e22}\,POT, %
with the beam in neutrino mode and the ND on-axis.



\begin{table}
	\centering
	\small
	\begin{tabular}{lccccc}
		\toprule
		&\textbf{PS191}	& \textbf{DUNE ND}& \textbf{SBND}	& \textbf{NA62} & \textbf{SHiP} \\
		\midrule
		Baseline& 128\,m		    & 574\,m			& 110\,m		    & 220\,m         & 60\,m          \\
		Volume  & 216\,m\tapi{3} & 150\,m\tapi{3} & 80\,m\tapi{3}  & 750\,m\tapi{3} & 590\,m\tapi{3} \\
		Energy	& 19.2\,GeV	    & 80\,GeV	    & 8\,GeV		    & 400\,GeV       & 400\,GeV       \\
		POT	    & \np{0.86e19}	& \np{1.32e22}	& \np{6.6e20}	& \np{3e18}     & \np{2e20}     \\
		\midrule
		Exposure& 1.0 	        & 220.9         & 16.4	        & 8.5           & 5820          \\
		\bottomrule
	\end{tabular}
	\footnotesize
	\caption{Comparison between experiments mentioned in this work.
		The exposure is defined as POT$\times$Energy$\times$Volume$\times$Baseline${}^{-2}$ with respect to %
		PS191, where ``Energy'' is the proton beam energy.
		The NA62 and SHiP experiments are not directly comparable with SBND and DUNE ND, %
		in that different technologies are involved;
		the RICH detectors are adopted as fiducial volume for NA62, whereas %
		for SHiP, we estimate the volume as the cone contained in the ``hidden sector'' vacuum vessel. 
		The volume is a driving feature in the definition of the total exposure and it is of utter importance %
		for searches of decay-in-flight events.}
	\label{tab:nd}
\end{table}

A summary of the features of the ND system is reported in~\reftab{tab:nd}, where it is compared to other beam dump experiments: %
PS191~\cite{Bernardi:1985ny,Bernardi:1987ek}, SBND which is the detector of the SBN programme with the best sensitivity to %
HNL~\cite{Ballett:2016opr}, NA62~\cite{NA62:2017rwk}, and SHiP~\cite{Anelli:2015pba}.
We define the total exposure of the experiment as the proton accelerator beam power, integrated over the total run time, %
and scaled by the volume of the detector over its baseline squared.
The beam power times the run time corresponds to the number of POT times the proton energy. 
With this definition, an exposure twelve times bigger is expected for the DUNE ND system with respect to SBND, %
and around two hundred times bigger than PS191.
The NA62 and SHiP experiments have a different design and are not directly comparable to TPC and tracker experiments, %
but we report them here for thoroughness.
The estimated exposure of NA62 is limited by its number of POT and by just one year of data taking; %
despite this fact, the experiment is optimised to study kaon decays and has good %
sensitivity to HNL~\cite{Drewes:2018irr}.
The SHiP experiment presents an exposure thirty times bigger than DUNE ND, but the detector is specifically %
designed to search for BSM physics, including heavy neutrinos~\cite{SHiP:2018xqw} (see also~\refref{Caputo:2016ojx}).
The decay-in-flight search hugely benefits from its 50\,m long decay vessel and short baseline.

On the collider physics frontier, the MATHUSLA~\cite{Curtin:2018mvb} and the FASER~\cite{Ariga:2018uku} experiments %
will perform a dedicated search for extremely weakly-interacting and long lived particles, %
like HNLs for which they presents interesting sensitivity~\cite{Curtin:2018mvb, Kling:2018wct}.
MATHUSLA will be a \np{800e3}\,m\tapi{3} hodoscope placed on the surface above the ATLAS or the CMS detectors.
FASER will consist of a 10\,m cylindrical decay volume located 480\,m downstream of the ATLAS interaction point. 

\subsection{Flux prediction}
\label{sec:tauneutrino}

\begin{figure}
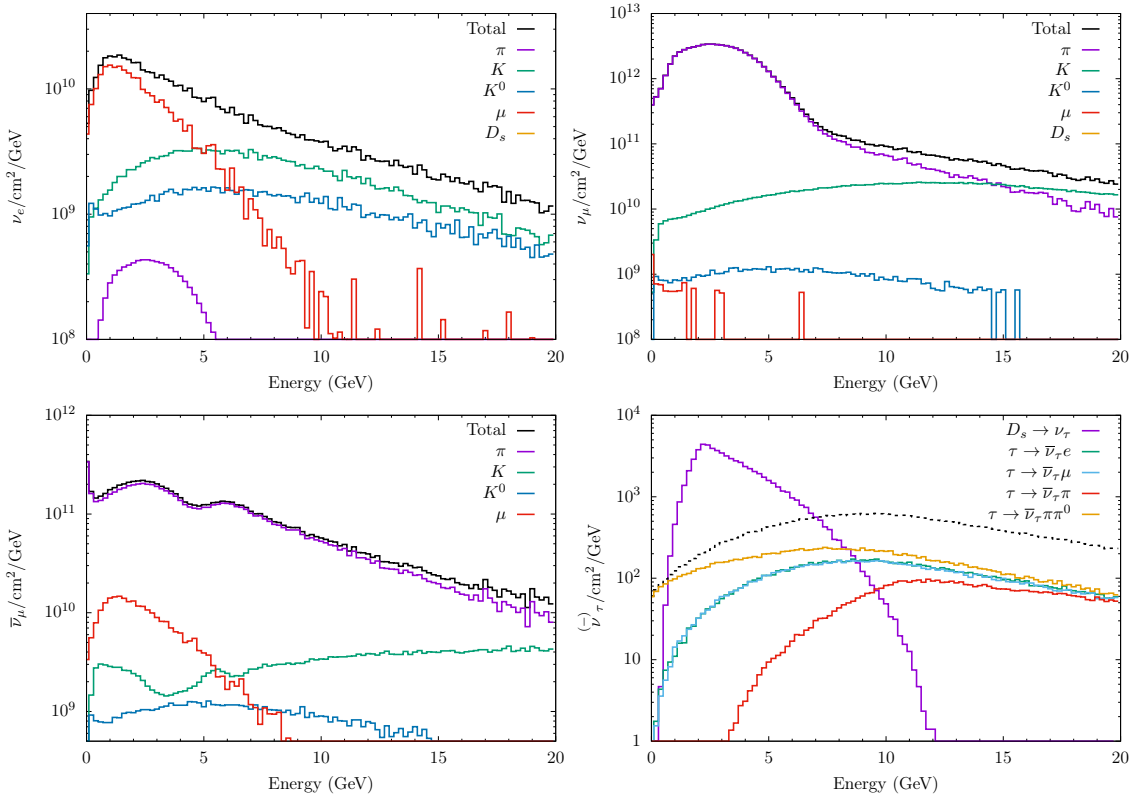

	\centering
	\resizebox{.5\textwidth}{!}{\input{fluxnue.tex}}
	\hspace{-1em}
	\resizebox{.5\textwidth}{!}{\input{fluxnumu.tex}}
	\\
	\resizebox{.5\textwidth}{!}{\input{fluxnubar.tex}}
	\hspace{-1em}
	\resizebox{.5\textwidth}{!}{\input{fluxnutau.tex}}
	\footnotesize
	\caption{The prediction of neutrino fluxes, in neutrino mode, divided by parentage at the ND are shown above, %
		normalised to \np{e20}\,POT.
		The $\nu_e$ component (top left) predominately originates from $\mu^+$ decays;
		kaon decays are responsible for the high energy part of the spectrum.
		The $\nu_\mu$ component (top right) obtains its main contribution from $\pi^+$ decays at low energies, %
		whereas the $K^+$ decays are accountable for the long tail of the spectrum.
		Contributions from $D_s^+$ decay are out of scale for both $\nu_e$ and $\nu_\mu$.
		The distribution of the $\cj{\nu}_\mu$ component (bottom left) is due to %
		negative charged secondary particles which are not successfully deflected by the horn system;
		the muon contribution is much more relevant than for the $\nu_\mu$ component.
		The $\nu_\tau$ component (bottom right) is only sourced from $D_s$ decays and presents a prominent peak at low energies, %
		whereas the $\cj{\nu}_\tau$ are produced in $\tau^+$ lepton decays.
		The dotted black line is the total $\cj{\nu}_\tau$ component of the flux.}
	\label{fig:fluxes}
\end{figure}

In order to implement our analysis, the various components of the flux by parentage must be known separately.
We study only the beam operating with a forward horn current, which selects %
positively charged secondary particles and results in a beam dominantly made of neutrinos %
with a smaller component of antineutrinos.
The flux predictions for $\nu_e$, $\nu_\mu$, and $\cj{\nu}_\mu$, provided by \refref{LauraFields} for the reference beam, %
are shown in~\reffig{fig:fluxes} subdivided in their parent components.
The $\nu_\mu$ flux is the dominant component and is principally originated %
by pion decays, whilst its long tail comes from kaon decays.
Unsuccessfully deflected negative particles, like $\pi^-$ or $K^-$, and the $\mu^+$ are the main contributors %
to the $\cj{\nu}_\mu$ components, and $\nu_e$ comes predominately from the muon %
and both $K^+$ and $K^0$ decays.
We consider only the energy range $E < 20\,\text{GeV}$, because it is the most intense region of the flux %
and, as it will be explained in~\refsec{sec:numevt}, the most relevant for this study.

\begin{figure}[t]
	\centering
	\resizebox{\linewidth}{!}{\input{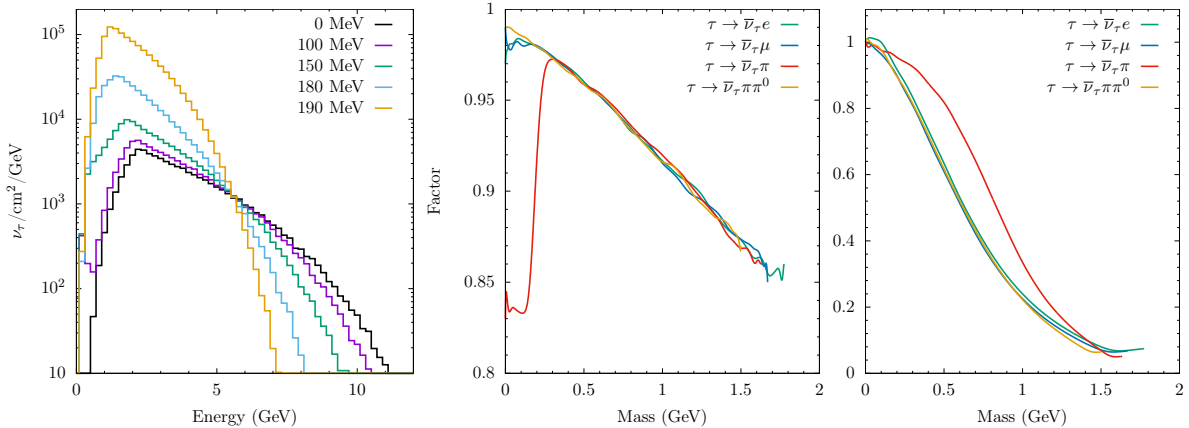}}
	\footnotesize
	\caption{The fluxes of heavy neutrinos from $D_s^+\to \tau^+ N$ (left) are presented %
		for different neutrino masses and normalised to \np{e20}~POT at the ND.
		Only phase space effects are considered here.
		For each different value of the neutrino mass, information on the start and end point of the spectrum %
		and the peak of the flux are extracted and used to reshape the $\nu_\tau$ spectrum.
		We show the distortion factors used in the scaling process for the channel producing $\cj{\nu}_\tau$: %
		the energy range normalised to 20\,GeV (middle) and the inverse of the peak re-scaling (right). }
	\label{fig:taudist}
\end{figure}

We highlight here the fact that we expect also an albeit-small flux of HNLs with masses above the kaon one.
This could be inferred from the $\nu_\tau$ flux, but this is not available in the literature.
In fact, the lightest meson with an interesting decay width to tau neutrinos is the charmed-strange meson~$D_s^+$, %
which has a mass $m_{D_s} = \np{1968.34}\pm\np{0.07}$\,MeV~\cite{PDG}.
It~decays into $\tau^+ \nu_\tau$ with a branching ratio of $\np{5.48}\pm\np{0.23}$\,\%~\cite{PDG}.
HNL with masses above the $K^0$ can be produced via the tau mixing, but more importantly via %
the muonic and electronic ones which are enhanced, as shown in \refsec{sec:production}.
The meson $D^+$ also decays into $\tau^+ \nu_\tau$, but being lighter than the $D_s^+$, %
the decay is disfavoured by the smaller phase space, with a branching ratio~50 times smaller.
This meson presents three-body decay channels into $\nu_e$ and $\nu_\mu$ with much higher branching ratio, %
but there is no enhancement for such channels into HNL, as explained in \refsec{sec:production}, and so these subdominant components %
are not taken in account in the present study.

The proton beam has a relatively low energy for producing charm quarks with a high cross-section, %
so the prediction of $\nu_\tau$ has not been carried out by the collaboration to the best of our knowledge.
For the reasons stated above, we make a prediction for the $D_s^+$ production by %
an 80\,GeV proton beam hitting a fixed graphite target.
The distribution at the production site will be then used to estimate the $\nu_\tau$ flux at the ND system.
In the literature, the following parametrisation has been successfully used to describe %
the charm meson production in proton-proton collision in the Centre of Mass frame~\cite{Ammar:1988ta}
\begin{equation}
	\label{eq:dsflux}
	\frac{\dd[2]{\sigma}}{\dd{x_F}\dd{p_T^2}} \sim (1-|x_F|)^n e^{-b p_T^2}~,
\end{equation}
where $x_F = 2 p_z/\sqrt{s}$, with $p_z$ the longitudinal momentum in the CM frame. %
The parameters $n$ and $b$ were fitted from the E769 experiment and found to be %
$n = \np{6.1}\pm\np{0.7}$ and \mbox{$b = \np{1.08}\pm\np{0.09}$}~\cite{Alves:1996qz}.
We assume that the $D_s^+$ meson production at the target follows the same distribution.
With the help of a Monte Carlo simulation, we generate the $D_s^+$ four-momenta starting from~\refeq{eq:dsflux} %
and simulate the meson decay and the subsequent tau decays.
A key simplification here is that because of the short lifetime of the $D_s^+$ and $\tau^+$, %
of the order of \np{e-13}\,s, their path is not affected by the horn system nor by interactions with other accelerator components. 
This results in no focusing of these secondary particles, and so only neutrinos emitted %
within the geometric acceptance of the ND are considered to form the $\nu_\tau$ and $\cj{\nu}_\tau$ spectrum.
The overall normalisation comes from an open charm calculation (see~\refapp{sec:opencc} for details): %
the number of $D_s^+$ per POT is found to be $(2.8 \pm 0.2) \times \np{e-6}$.
The result of the simulation is reported in the bottom right panel of~\reffig{fig:fluxes}, %
where the different contributions to the $\nu_\tau$ spectrum are shown.
Thanks to the large number of POTs in DUNE, the total number of $D_s^+$ mesons produced is comparable %
to other dedicated experiments~\cite{Alekhin:2015byh}; %
however, the beamline design is not optimised for heavy mesons production %
and the $\nu_\tau$ flux seen at the ND is strongly attenuated.

Having knowledge of the parent meson distribution, we directly simulate the production of nearly-sterile neutrinos %
from the $D_s$ decays.
The spectrum of heavy neutrinos is distorted when their mass approaches the various phase space thresholds, %
which appears as a further enhancement of the flux. 
This is because heavier neutrinos are more easily boosted inside the geometric acceptance of the detector.
Besides the peak height, the start and the end point of the energy flux are also affected,
as illustrated in~\reffig{fig:taudist}. 
We take these effects in account, modifying the scaled neutrino flux using information retrieved by the $\nu_\tau$ and $\cj{\nu}_\tau$~simulation. 

\subsection{Background evaluation}
\label{sec:background}


\begin{table}
	\newcommand{\us}{\hphantom{${}^0$}}
	\newcommand{\ms}{\hphantom{${}^-$}}
	\centering
	\small
	\begin{tabular}{lrrr@{\,}lrrr@{\,}l} 
		\toprule
		& \multicolumn{3}{c}{CC events}	&  \multicolumn{3}{c}{NC events}	\\
		\cmidrule(lr){2-5} \cmidrule(lr){6-9}
		& Per tonne	& Ratio		& \multicolumn{2}{c}{Rate (Hz)}	& Per tonne 	& Ratio	& \multicolumn{2}{c}{Rate (Hz)}	\\
		\cmidrule(lr){2-5} \cmidrule(lr){6-9}
	$\nu_e$		    & %
	\np{3.0e3}\ms	& 75.6\,\%	& 152  & $\times\,10^{-3}$ & \np{1.0e3}\ms  & 24.4\,\% & 48.9 & $\times\,10^{-3}$	\\
	$\nu_\mu$	    & %
	\np{236e3}\ms	& 75.2\,\%	& 12.0 &                 & \np{77.8e3}\ms & 24.8\,\% & 3.95 &                  	\\
	$\cj{\nu_\mu}$	& %
	\np{17.7e3}\ms& 70.9\,\%	& 898  & $\times\,10^{-3}$ & \np{7.2e3}\ms  & 29.1\,\% & 368 & $\times\,10^{-3}$	\\
	$\nu_\tau$	    & %
	\np{1.6e-5}	& 17.1\,\%      & 8.3  & $\times\,10^{-10}$& \np{7.9e-5}    & 82.9\,\% & 4.0 & $\times\,10^{-10}$	\\
	$\cj{\nu_\tau}$	& %
	\np{5.2e-5}	& 45.3\,\%      & 2.6  & $\times\,10^{-3}$ & \np{6.1e-5}    & 54.7\,\% & 3.0 & $\times\,10^{-9}$	\\
		\bottomrule
	\end{tabular}
	\footnotesize
	\caption{The expected rates for CC and NC interaction in the near detector are presented here, normalised to %
		\np{e20}\,POT.
		The values were computed starting from~\refeq{eq:numev}, convolving the fluxes of~\reffig{fig:fluxes} 
		with the CC and NC cross section predictions from GENIE~\cite{Andreopoulos:2009rq}.
		Detector efficiencies are not applied.
		The first columns show the total number of events per tonne of argon, the second ones %
		the proportion of CC or NC events with respect to the totality, and the last columns the event frequencies %
		assuming \np{1e14}\,POT/s.}
	\label{tab:rate}
\end{table}

The number of SM neutrino--nucleon interactions expected at the DUNE ND, without considering detector effects, is calculated %
by integrating the Charged Current (CC) and Neutral Current (NC) total cross sections multiplied %
by the light neutrino spectrum~$\dv*{\phi_\nu}{E}$:
\begin{equation}
	\label{eq:numev}
	\mathcal{N}_\text{tot} = \mathcal{N}_\text{CC} + \mathcal{N}_\text{NC} = 
	\mathcal{N}_\text{target} \int \dd{E} \qty[\sigma_\text{CC}(E) + \sigma_\text{NC}(E)] \, \dv{\phi_\nu}{E}\ ,
\end{equation}
where $\sigma_\text{CC}(E)$ and $\sigma_\text{NC}(E)$ are the cross section predictions in argon %
calculated with \mbox{GENIE}~\cite{Andreopoulos:2009rq}, and $\mathcal{N}_\text{target}$ is the %
total number of Ar targets. 
The event rates are shown in~\reftab{tab:rate}.
It turns out that less than one $\nu_\tau$ event is expected in the total run of the experiment.
As~a~comparison, the number of $\nu_\mu$ events will be \np{e10} times higher.
This confirms the expectations that the $\nu_\tau$ component of the flux is negligible %
for standard oscillation physics in~DUNE ND.
On the other hand $\nu_\tau$ appearance is expected at the FD.


These neutrino scatterings occurring within the fiducial volume of the detector could mimic %
the rare signal of neutrino in-flight decays, as some final state particles are common to both processes.
A good estimate of the number of possible background events for each discovery channel is very important, %
since it dictates the true sensitivity of the experiment.
We restrict our conservative background analysis to decay modes available for neutrino masses below $m_{K^0}$, %
being these the channels with the best discovery potential.
They are $N\ra\nu e^+ e^-$, $\nu e^\pm \mu^\mp$, $\nu \mu^+ \mu^-$, $\nu \pi^0$, $e^\mp \pi^\pm$, and $\mu^\mp \pi^\pm$.
Particles are typically tagged by studying the topology of the tracks and the energy loss $\dv*{E}{x}$ in the active medium, %
but instead of dealing with a full detector simulation, we perform a fast Monte Carlo analysis, %
using as input neutrino--nucleon scattering events in argon generated by the neutrino event generator GENIE~\cite{Andreopoulos:2009rq}.
The tracks are randomly placed inside the ND system and then smeared according to a normal distribution centred on the simulated value of energy/momentum; %
particles with a kinetic energy above the detection threshold are then assumed to be reconstructed.
The relative position between the two detectors is taken into account, in that %
particle tracks exiting the LArTPC end entering the MPD are reconstructed as a single track.
Escaping or partially reconstructed tracks are not discarded, but treated with a different energy/momentum resolution: %
the initial particle energy can be estimated, with some limitations, thanks to the energy dependence of the mean energy loss %
during the particle propagation.
We then implement possible sources of background mis-identification which are channel specific.
Detector resolutions and thresholds, from~\refref{Alion:2016uaj} for both parts of the ND, are summarised in~\reftab{tab:fastmc}.


\begin{table}
	\centering
	\small
	\begin{tabular}{lccc}
		\toprule
		Particle& Threshold	& $\sigma_\text{rel}$	&  $\sigma_\theta$		\\
		\midrule
		EM	& 30\,MeV	& $5\%/\sqrt{E} \oplus 1\%$	& 1\textdegree	\\
		Hadron	& 50\,MeV	& $30\%/\sqrt{E} \oplus 5\%$	& 5\textdegree	\\
		Muon	& 30\,MeV	& 1\% or 30\% of $|\vb{p}|$	& 0.3\textdegree	\\
		Pion	& 100\,MeV	& 1\% or 30\% of $|\vb{p}|$	& 0.3\textdegree	\\
		\bottomrule
	\end{tabular}
	\footnotesize
	\caption{The table lists detection thresholds and energy/momentum and angular resolutions used in the fast MC, %
		where ``EM'' delineates electro-magnetic showers and ``Hadron'' any other charged particle %
		which is neither a lepton nor a pion.
		The momenta of pions and muons are smeared according to the containment of their tracks.
		If the particles enter the MPD in which they cover a length longer than the detector's diameter or %
		if 80\,\% of the tracks are contained inside the LArTPC then the relative resolution on the momentum is 5\,\%, %
		otherwise a resolution of 30\,\% is applied.
		Neutrons are treated with ``Hadron'' resolutions, but with a 90\,\% detection efficiency. }
	\label{tab:fastmc}
\end{table}
A strong discriminant for background events is the presence of protons, neutrons, and other hadrons in the final states, %
which are the results of the nucleus recoil of the neutrino interaction.
If hadronic activity is reconstructed as an interaction vertex, then the event is clearly originated by %
SM neutrino--nucleon scattering and tagged as background.
In the case this does not happen, for instance when the hadrons are below threshold, the multiplicity of final state particles %
becomes fundamental to distinguish signal events from intrinsic background.
However, this background can be worsened by mis-identification of certain tracks.

The main background to the pseudo-scalar meson channels, $N\to \ell^\mp \pi^\pm$, are resonance $\nu_e$ or $\nu_\mu$--CC %
interaction with single pion production or charged current incoherent and deep inelastic scatterings %
in which only a pair $\ell\,\pi$ is detected.
Three-body lepton decays suffers from mis-identification of additional pions and photons emitted in CC neutrino scatterings %
which are mistaken for charged leptons.
Despite having a similar mass, pion and muon tracks differ on average in length, as the meson track often culminates in a hadronic shower.
In~our implementation of the detector effects, if no hadronic shower is detected and the track length is longer than two metres, %
the pion is identified as a muon.
Electromagnetic shower induced by photons are identified by looking at the vertex displacement and at the $\dv*{E}{x}$, %
which is twice as large as the energy loss for $e^\pm$.
If a photon converts within two centimetres from the interaction point, and either the electron or the positron of the pair is below threshold, %
the photon is reconstructed as a single electron.
A pair of electrons with a small separation angle, less than 3\textdegree, is tagged as an electron-positron pair %
and the parent photon is reconstructed.
The main source of photons comes from the decay of the neutral pion, which is abundantly produced in %
NC neutrino--nucleon interactions.
Certain hadronic transitions from secondary particles of deep inelastic scatterings also emit gammas.
If a pair of photons shows an invariant mass comparable with the $\pi^0$ mass, the parent pion is identified.
Interactions in which multiple neutral pions are produced, but only a pair of photons is detected and reconstructed, %
are background to the $N\to \nu \pi^0$ channel.
The~background events surviving particle identification are between 2.5\,\% down to 0.0025\,\% of the processed events.


The channels which open up for masses above the kaon mass are more challenging from an experimental point of view.
The final state particles of these modes are mostly neutral pseudo-scalar mesons, which decay electromagnetically, %
or vector mesons, which usually decay into a multi-state of lighter mesons, depending on the initial flavour content,
sometimes accompanied by photon emission.
The correct identification of these short-lived states is non trivial.
For very high masses, also $\tau$ leptons are yielded, but their precise reconstruction requires \emph{ad hoc} techniques.
These tasks are beyond the scope of the analysis presented here and are best left to the collaboration superior simulation tools.
We also do not consider cosmogenic background, even though a rate of \np{2.7}\,Hz/m\tapi{2} cosmic rays %
is expected at the ND hall~\cite{Sinclair:DUNEdoc}, which has very little over burden.
Given an area of a few square meters, the number of cosmic rays per drift window can be non-negligible~\cite{Abi:2018dnh}, %
but rejection techniques are being developed with good signal efficiencies~\cite{Adams:2018lzd}.

%
\subsection{HNL decay events and signal efficiency}
\label{sec:numevt}

Except for $N$ decaying into three neutrinos, all the other decay channels are in principle detectable.
For a given visible decay mode~$d$, the number of signal events is
\begin{equation}
	\label{eq:event}
	\mathcal{N}_d = \int\!\! \dd{E}\ \Pi_d(E) W_d(E)\, \dv{\phi_N}{E}\ ,
\end{equation}
where $\dv*{\phi_N}{E}$ is the number of heavy neutrinos expected at the ND, %
computed in the way described in~\refsec{sec:production}.
The function $\Pi_d(E)$ accounts for the probability of a heavy neutrino of energy $E$ to decay inside the ND after covering the baseline distance $L$.
It is expressed in the following form:
\begin{equation}
	\Pi_d(E) = e^{-\frac{\Gamma_\text{tot} L}{\gamma \beta}} %
	\qty(1-e^{-\frac{\Gamma_\text{tot} \lambda}{\gamma \beta}}) \frac{\Gamma_d}{\Gamma_\text{tot}}\ , 
\end{equation}
where $\lambda$ is the length of the ND, $\Gamma_d$ the decay rate for the channel $d$ and %
$\Gamma_\text{tot}$ the total decay rate.
The total effect of $\Pi_d$ is to favour low-energy bins of the neutrino spectrum for which the %
relativistic factor $\gamma\beta$ is small.

The term $W_d(E)$ is a signal efficiency factor, estimated as the binned ratio of the true $N$ energy spectrum after %
and before a background rejection procedure.
This process aims at further reducing the number of background events still present after particle identification.
It consists of simple data selection cuts optimised to reject the background while keeping an acceptable signal efficiency %
(typically above 30\,\%), exploiting differences in the energy and angular distributions between signal and background events.
The HNL decays inside the detector are simulated by a custom Monte Carlo code and the tracks are processed in the same way %
as it is done for background events (see~\refsec{sec:background}).
The resulting signal efficiency therefore embeds also detector effects.
If no background is expected for the channel $d$, there is no need for applying any rejection procedure %
and so the signal efficiency is maximal, i.e.\ $W_d(E) = 1$ at all energies.
The final number of background events $\mathcal{B}_d$ and the number of signal events $\mathcal{N}_d$ are %
eventually used to build the Confidence Level (C.L.) regions of sensitivity (see \refsec{sec:results}).
%
We leave a more detailed discussion on the background reduction cuts in~\refapp{sec:appbackground}, %
where we report the rates of background reduction and signal selection for all decay channels of both Majorana and Dirac neutrinos of a given mass.
From our analysis, we note that selection cuts are slightly different for Dirac or Majorana HNL decays.
This is a consequence of certain combinations of production and decay modes which are forbidden for Dirac neutrinos, %
as they would lead to LNV, and so the energy and angular distributions are not identical.
NC--mediated decay mode have also intrinsically distinct decay widths in the heavy neutrino rest frame %
and the difference angular dependency can be reflected in the laboratory frame.

\section{Sensitivities of DUNE ND}
\label{sec:results}

We present here sensitivity regions for the discovery of heavy neutrino decays for %
a total amount of \np{1.32e22}\,POT collected with the beam in neutrino mode.
All the regions are estimated at the 90\,\% C.L.\ in rejecting the null hypothesis, %
by which no HNL decays are seen ($\sigma = 0$), but only background events $b$ are expected.
For a specific channel~$d$, the probability of observing $n$ events with a signal mean $\sigma = \mathcal{N}_d$ %
and background $b = \mathcal{B}_d$ (see~\refsec{sec:numevt}) follows a Poisson distribution
\begin{equation*}
	P(n|\sigma,b) = (\sigma+b)^n \frac{e^{-(\sigma+b)}}{n!}\ .
\end{equation*}
We employ the Feldman and Cousins method~\cite{Feldman:1997qc} to estimate the number of events needed in order %
to reject $H_0$ at the desired C.L..
For example, if no background is expected \mbox{($W_d = 1$)}, an average of $n = 2.44$ events %
must be detected to reject $H_0$ with 90\,\%~C.L.\ 
This~criterion is used to define the sensitivity regions shown in this section, for both Majorana and Dirac neutrinos.
It is expected that the MPD alone has a better sensitivity than the LArTPC, %
thanks not only to a larger volume, but also to a less dense medium which gives lower backgrounds.
As the two modules are assumed to have the same detection performance, we present here a combined analysis of the %
two detectors, taking into account particle propagation between them.	
We do not consider charged identification capabilities of the ND, and therefore this information is washed out %
in presenting the sensitivity plots in this and next sections.
Because of our charge-blind analysis, the number of events expected for Majorana neutrinos is twice as large as %
the number in the case of Dirac neutrinos, therefore the sensitivity to Dirac neutrino decays is %
a factor of $\sqrt{2}$ worse than the Majorana case.\footnote{The sensitivity for high number can be roughly estimated as %
$\flatfrac{\mathcal{N}_d}{\sqrt{\mathcal{N}_d + \mathcal{B}_d}}$, %
and for zero background it simply scales as $\sqrt{\mathcal{N}_d}$.}
The limits reported here below refer to Majorana heavy neutrinos; %
the corresponding limit for which $N$ is a Dirac fermion is easily retrieved by multiplying the upper limit %
by $\sqrt{2}$.

In~\refsec{sec:dominant}, we show the constraint that DUNE ND can place on a simplified scenario %
in which a single mixing matrix element between HNL and active neutrinos dominates.
We~have also considered a scenario in which two mixings are dominant with respect to the third one, %
the results of which are presented in~\refsec{sec:bimax}.

\subsection{Single dominant mixing}
\label{sec:dominant}

\begin{figure}
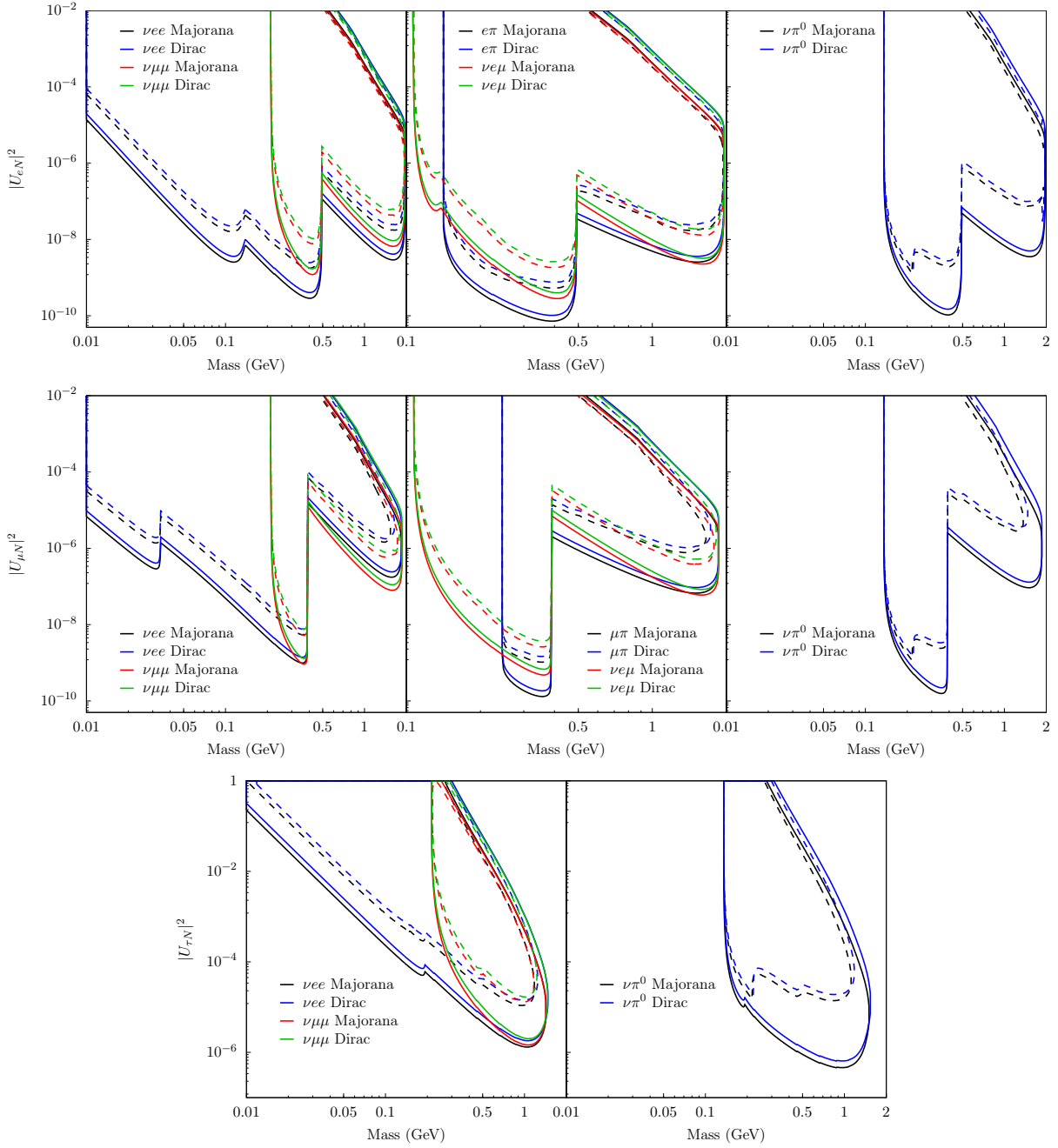

	\centering
	{\resizebox{\linewidth}{!}{\input{sensmulti_EW2_real.tex}}}
	\vspace{0.05em}

	{\resizebox{\linewidth}{!}{\input{sensmulti_MW2_real.tex}}}
	\vspace{0.05em}

	{\resizebox{\linewidth}{!}{\input{sensmulti_TW2_real.tex}}}

	\footnotesize
	\caption{The 90\,\% C.L. sensitivity regions to individual channels for dominant mixings %
		$|U_{e N}|^2$ (top), $|U_{\mu N}|^2$ (middle), and $|U_{\tau N}|^2$ (bottom) are shown.
		The solid lines correspond to the analysis before the background analysis, which is equivalent %
		to a weighting factor $W_d = 1$ (see~\refeq{eq:event}).
		The dashed lines are drawn after our background analysis.
		The distinction between the fermionic natures are explained in the colour key.}
	\label{fig:senseW}
\end{figure}

In this section, we present the sensitivity regions for the three mixings $|U_{e N}|^2$, $|U_{\mu N}|^2$, and %
$|U_{\tau N}|^2$, where we assume that just one mixing element dominates over the other two.
The~sensitivities for the decay channels $N\ra\nu e^+ e^-$, $\nu e^\pm \mu^\mp$, $\nu \mu^+ \mu^-$, $\nu \pi^0$, %
$e^\mp \pi^\pm$ ($|U_{e N}|^2$ only), and $\mu^\mp \pi^\pm$ ($|U_{\mu N}|^2$ only) are reported in \reffig{fig:senseW}.
The solid lines corresponds to a scenario in which zero background is assumed at the ND.
A background study is done for these channels (see~\refsec{sec:background}), to outline a more realistic sensitivity; %
the resulting regions are shown as dashed lines in \reffig{fig:senseW}.
As we expect that further improvements to background reduction can be achieved %
with a dedicated analysis by the experimental collaboration, %
the final sensitivity will lie somewhere between the lines with and without backgrounds.

For both the electronic and the muonic mixings, the two-body semi-leptonic decay modes are the ones providing %
the best sensitivity for sufficiently heavy masses.
With the channel $N\to e^\mp \pi^\pm$, the mixing can be constrained in the range %
$0.15\,\text{GeV} \lesssim m_N \lesssim 0.49\,\text{GeV}$ %
to be $|U_{e N}|^2 < \np{3e-9}$, with a minimum point $|U_{e N}|^2 < \np{7e-11}$ at $m_N \simeq 0.39$\,GeV.
Including the background rejection, the limits are loosened by a factor of $\sim$6.1.
The~channel $N\to \mu^\mp \pi^\pm$ can constrain the mixing $|U_{\mu N}|^2 < \np{5.6e-10}$ %
in the mass range \mbox{$0.25\,\text{GeV} \lesssim m_N \lesssim 0.39\,\text{GeV}$}, %
with the best limit $|U_{\mu N}|^2 < \np{1.3e-10}$ at $m_N \simeq 0.35$\,GeV.
In this case, the higher background reduce the bounds up to a factor of $\sim$14.3.
The NC decay $N\to \nu \pi^0$ is the channel most affected by background and with the worst signal efficiency: %
the limits are higher at most by a factor of~$\sim$29.6 for the electronic, %
$\sim$36.5 for the muonic, and~$\sim$42.5 for the tau mixing.
Assuming no background, instead, the constrains placed by this decay mode can be competitive, as the %
mixings are limited to be $|U_{e N}|^2 < \np{1.1e-10}$ at $m_N \simeq 0.39$\,GeV, %
$|U_{\mu N}|^2 < \np{1.5e-10}$ at $m_N \simeq 0.35$\,GeV, %
and $|U_{\tau N}|^2 < \np{6.70e-7}$ at $m_N \simeq 0.95$\,GeV.
There is no sensitivity to the channel $N\ra\tau^\mp\pi^\pm$ because of the subdominant branching ratio %
and flux content.

The three-body lepton decays have a lower reach, but are more sensitive to masses above the kaon mass limit %
than the two-body semi-leptonic modes.
The channel $N\to \nu e^- e^+$ is the only one that covers the whole mass range of interest %
and the bounds are weakened by background reduction by a factor less than 6.
It can limit the electronic mixing down to $|U_{e N}|^2 < \np{2.5e-9}$ at $m_N \simeq 0.11$\,GeV, %
$|U_{e N}|^2 < \np{2.9e-10}$ at $m_N \simeq 0.39$\,GeV, and $|U_{e N}|^2 < \np{3.0e-9}$ at $m_N \simeq 1.6$\,GeV.
The channels $N \to\nu \mu^- \mu^+$ and $\nu e^\pm \mu^\mp$ perform better with the muon mixing, %
despite suffering more from background rejection, up to a factor of 16 for the muon mixing and a factor of 17 %
for the tau mixing.
They respectively give the limits %
$|U_{\mu N}|^2 < \np{9.0e-10}$ at $m_N \simeq 0.37$\,GeV and $|U_{\mu N}|^2 < \np{8.2e-8}$ at $m_N \simeq 1.6$\,GeV, and %
$|U_{\mu N}|^2 < \np{4.7e-10}$ at $m_N \simeq 0.36$\,GeV and $|U_{\mu N}|^2 < \np{6.1e-8}$ at $m_N \simeq 1.6$\,GeV.
The $\tau$ sector can only be constrained by the two NC--mediated channels, %
which give very similar constraints near $m_N\simeq 1.0$\,GeV, these being $|U_{\tau N}|^2 < \np{2.2e-6}$ 
for the $\nu e^- e^+$ channel and $|U_{\tau N}|^2 < \np{2.2e-6}$ for the $\nu \mu^- \mu ^+$ channel.


\begin{figure}
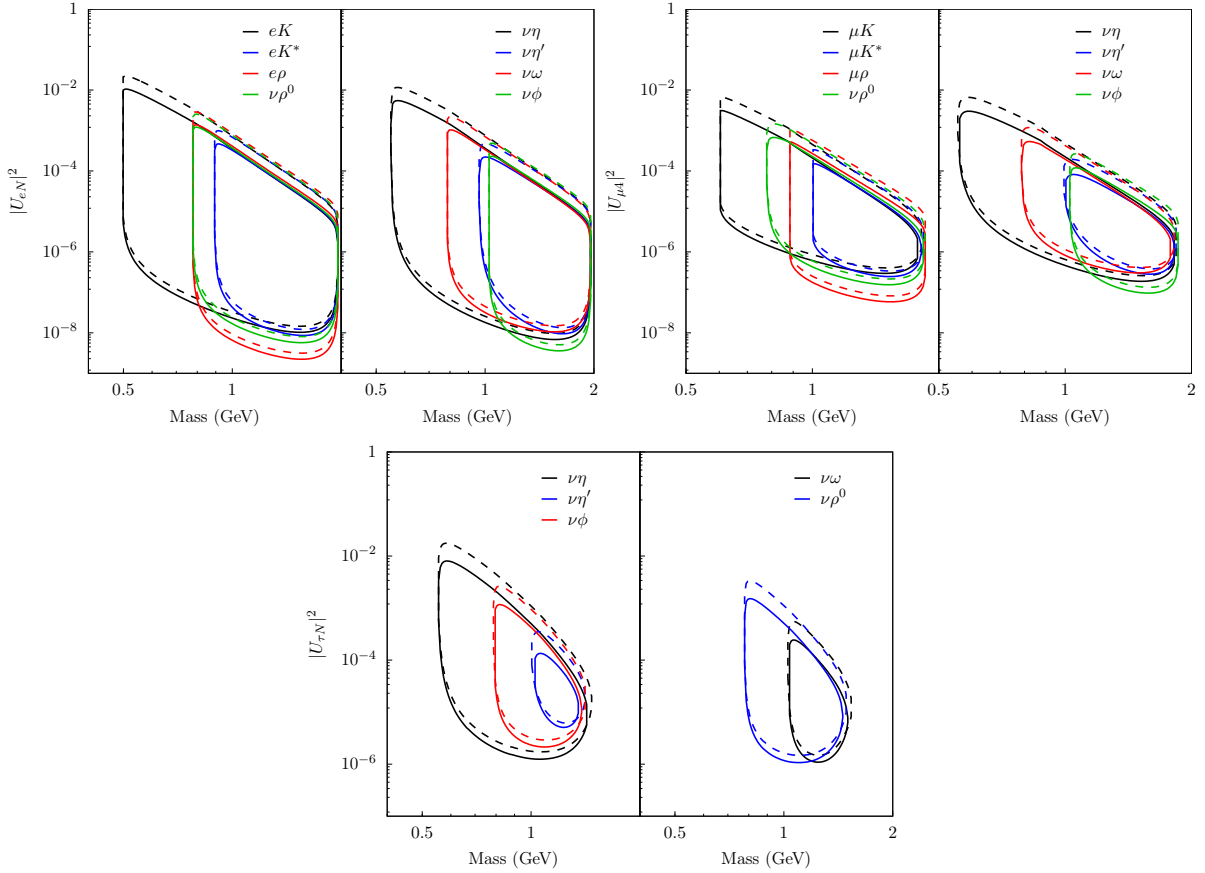

	\centering
	{\resizebox{\linewidth}{!}{\input{sensmulti_EM_V2_real.tex}}}
	\vspace{0.05em}

	{\resizebox{\linewidth}{!}{\input{sensmulti_T_V2_real.tex}}}
	\footnotesize
	\caption{The 90\,\% C.L. sensitivity regions to individual channels for dominant mixings %
		$|U_{e N}|^2$ (top left), $|U_{\mu N}|^2$ (top right), and $|U_{\mu N}|^2$ (bottom) are presented for Majorana (solid lines) %
		and Dirac (dashed lines) neutrinos.
		No background analysis was performed for the channels shown here (see text).
		These channels become available only for masses above 0.5\,GeV.
	}
	\label{fig:senseV}
\end{figure}

A background study was not performed for all the other decay channels, which open up for masses above the $K^0$ mass, %
due to the fact that the final state particles need a more complex analysis.
The sensitivities to these modes are shown in~\reffig{fig:senseV}, and they can place some constraints to the mixing.
All the channels peak in their sensitivity for masses between $1.3$ and $1.8$\,GeV.
The best limits obtained for CC decays are %
$|U_{e N}|^2 < \np{2.3e-9}$ from $N \ra e^\mp \rho^\pm$ and $|U_{\mu N}|^2 < \np{6.0e-8}$ from $N \ra \mu^\mp \rho^\pm$; %
among the NC decays $|U_{e N}|^2 < \np{3.7e-9}$ and $|U_{\mu N}|^2 < \np{1.0e-7}$ both from $N \ra \nu \phi$.
Even for these channels, there is no sensitivity to CC processes to the tau mixing, %
but interesting limits are set from $N\to \nu\eta$, $N\to \nu\omega$, and $\nu\rho^0$ to be respectively %
$|U_{\tau N}|^2 < \np{1.86e-6}$, $\np{3.24e-6}$, and~$\np{1.60e-6}$.

\subsection{Two dominant mixings}
\label{sec:bimax}

\begin{figure}
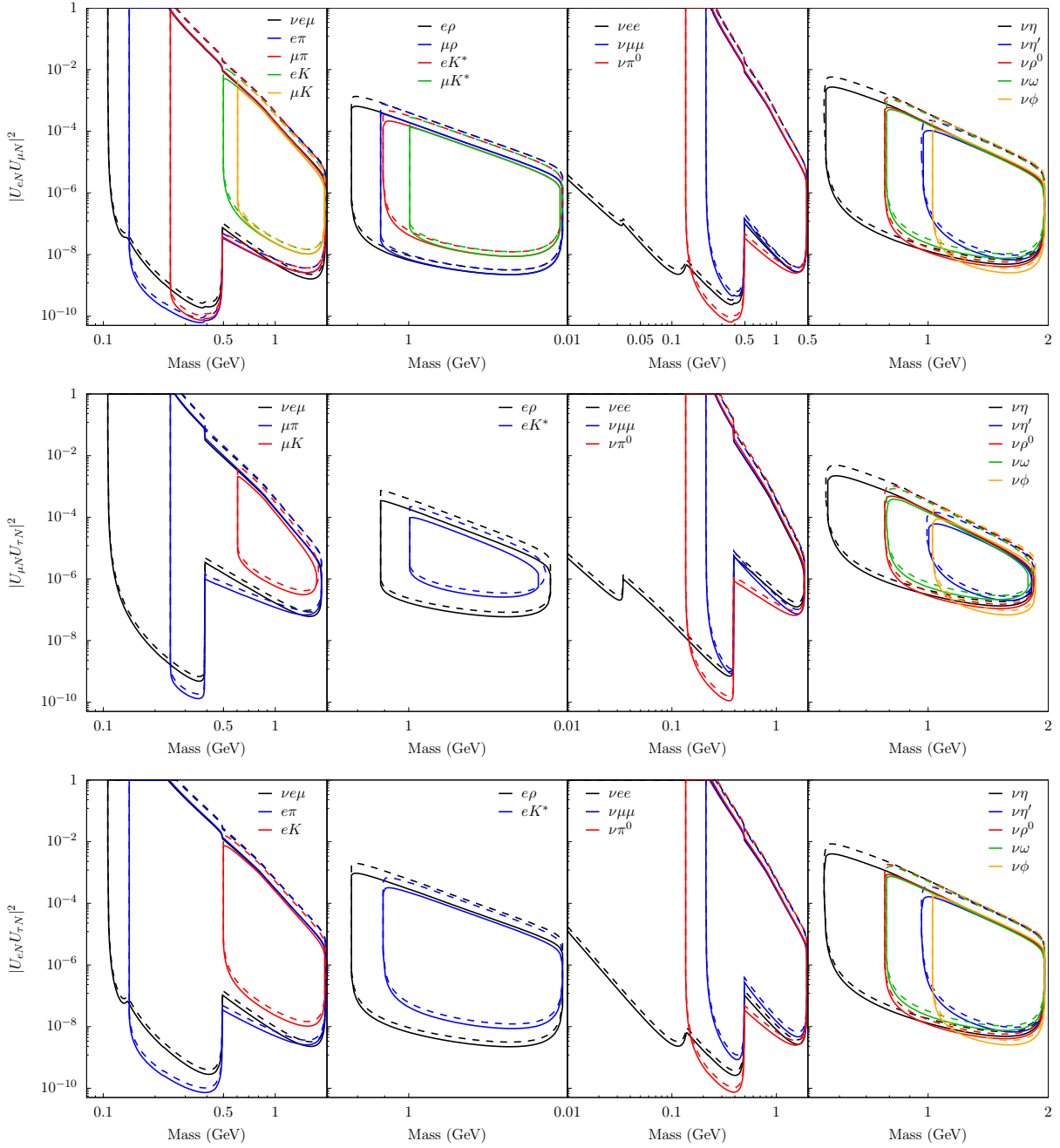

	\centering
	{\resizebox{\linewidth}{!}{\input{sensmulti_EM2_real.tex}}}
	\vspace{0.05em}

	{\resizebox{\linewidth}{!}{\input{sensmulti_MT2_real.tex}}}
	\vspace{0.05em}

	{\resizebox{\linewidth}{!}{\input{sensmulti_TE2_real.tex}}}
	\footnotesize
	\caption{The 90\,\% C.L. sensitivity regions to individual channels for two dominant mixings %
		$|U_{e N}^* U_{\mu N}|$ (top), $|U_{\mu N}^* U_{\tau N}|$ (middle), and $|U_{e N}^* U_{\tau N}|$ (bottom) are presented.
		All the modes considered in this work are shown here, but no background analysis is reported.
		As before, the solid lines correspond to the analysis with Majorana neutrinos, the dashed lines with Dirac neutrino.}
	\label{fig:senseMix}
\end{figure}

In this section we present the bounds in a scenario in which two mixing elements are comparable and dominant over the third one.
This case complements the previous analysis in \refsec{sec:dominant} as, by searching for HNL decays, %
the experiment can constrain certain combinations of the mixing elements.
This can happen when the neutrino is produced via one mixing and decays via another one, %
or when both mixing elements play a role in production and decay.
For instance, the decay $K^+ \ra \mu^+ N$ yields heavy neutrinos with a flux proportional to %
$|U_{\mu N}|^2$, but they can afterwards decay into the channel $\nu e^+ e^-$ also via the electronic or the tau mixing.
It is important to highlight that, in the case in which one mixing is responsible %
for the production and a different mixing for the decay, %
then number of events is proportional to the product of the mixings %
$|U_{\alpha N}||U_{\beta N}|$ if the studied channel is CC--mediated.
However, if the decay channel is also sensitive to a NC exchange, the number of events is instead proportional to %
$|U_{\alpha N}|\sqrt{|U_{\alpha N}|^2 + |U_{\beta N}|^2}$.
In the remainder of this section, we will use the combination of two mixings represented by $|U_{\alpha N}^* U_{\beta N}|$ %
\enlargethispage{\baselineskip}
for comparing bounds and sensitivity plots.

The combinations of mixing terms is relevant to charged Lepton Flavour Violating (cLFV) %
decays or flavour changing neutral current processes %
which can be enhanced in presence of nearly-sterile neutrinos.
For example, the well-known decay $\mu^+ \to e^+ \gamma$ has a branching ratio which is sensitive to extra neutrino states.
This reads
\begin{equation}
	\label{eq:meg}
	\text{Br}(\mu^+ \ra e^+ \gamma) = \frac{3 \alpha}{32 \pi} \abs{\sum_i \hat{U}_{\mu i}^*\, \hat{U}_{e i}\,G\qty(\frac{m_i^2}{M_W^2})} \ ,
\end{equation}
where $G(x)$ is the loop function of the process~\cite{Ilakovac:1994kj}.
The current upper limit is set by the MEG experiment to be %
$\text{Br}(\mu^+ \ra e^+\gamma) < \np{4.2e-13}$~\cite{TheMEG:2016wtm}.
Despite being one of the best constrained cLFV process, the bounds on $|U_{e N}^* U_{\mu N}|$ are not as good as the ones imposed %
by other processes, like $\mu \to e e e$ or $\mu - e$ conversion on nuclei~\cite{Alonso:2012ji}.
For instance, the constraint from conversion on Au is $|U_{e N}^* U_{\mu N}| < \np{1.6e-5}$ %
for HNL masses larger than 0.1\,GeV~\cite{Deshpande:2011uv}.
The branching ratio of other cLFV channels, like $\tau \to e \gamma$ or $\tau \to \mu \gamma$ are not as well constrained %
and so the bounds achievable on the combination of heavy neutrino mixings are expected to be less %
stringent~\cite{Buras:2010cp, Abada:2016vzu}.
Stronger bounds come from study of three-body decays of charm and bottom mesons to charged leptons with different flavour %
and tau decays to pseudo-scalar mesons and a charged lepton: from the search for %
the decay $K \to e \mu \pi$ the bound $|U_{e N}^* U_{\mu N}| < \np{e-9}$ is reached for masses %
$0.15\,\text{GeV}\lesssim m_N \lesssim 0.50\,\text{GeV}$;
the decays $\tau \to e \pi \pi$ and $\tau \to \mu \pi\pi$ set the limits %
$|U_{e N}^* U_{\tau N}|, |U_{\mu N}^* U_{\tau N}| < \np{5e-6}$ for the respective mass ranges %
$0.14\,\text{GeV}\lesssim m_N \lesssim 1.7\,\text{GeV}$ and %
$0.24\,\text{GeV}\lesssim m_N \lesssim 1.7\,\text{GeV}$~\cite{Helo:2010cw}.

Instead of dealing with a three-dimensional parameter scan, we simplify the study %
by assigning the same value to the two mixing parameters under consideration, for which the number %
of HNL decays is maximal.
The number of events is then reported as a function of the neutrino mass and the combination $|U_{\alpha N}^* U_{\beta N}|$.
The results for all channels considered in this work are shown in~\reffig{fig:senseMix}.
The best constraints come again from two-body semi-leptonic decays for all mixing combinations, %
the lowest upper limits being $|U_{e N}^* U_{\mu N}| < \np{6e-11}$ at $m_N \simeq 0.36$\,GeV, %
$|U_{\mu N}^* U_{\tau N}| < \np{1.3e-10}$ at $m_N \simeq 0.35$\,GeV, %
and $|U_{\tau N}^* U_{e N}| < \np{7e-11}$ at $m_N \simeq 0.39$\,GeV.
Amongst the three-body leptonic decay channels, $N\to\nu e e$ has the best sensitivity for masses $m_N < m_{K^0}$, %
but actually the mode $N\to \nu e^\mp \mu^\pm$ can be more constraining at higher masses.
Regarding the channels available only above the kaon mass threshold, decays to pseudo-scalar mesons are the most sensitive %
between CC processes, whereas the decay $N \to \nu \phi$ gives the best constraint of the NC--mediated channels.

\section{Mass model constraints from DUNE ND}
\label{sec:combined}


From the results presented in the previous section, we find that the DUNE ND will be sensitive to very low couplings %
for experimentally accessible mass values.
These points of the parameter space corresponds to regions viable in some realisations of low scale neutrino mass models.
In view of the discussion regarding seesaw models in \refsec{sec:model}, we perform a mass matrix random scan to %
define such regions of the parameter space.
Following the previously introduced notation, we focus on three minimal ISS scenarios which predict a HNL with a mass accessible %
by the experiment and that satisfy the experimental evidence of neutrino oscillation~\cite{Abada:2014vea}.
In the first two cases, the heavy neutrino under study belongs to the lightest pseudo-Dirac pair %
of an ISS\,(2,2) and an ISS\,(2,3) realisation; the third scenario is an ISS\,(2,3) case %
in which the fourth Weyl state becomes a Majorana neutrino in the \mbox{MeV--GeV} region thanks to a high LNV parameter.
The details of this analysis are reported in this section, together with the overall sensitivities of DUNE ND to %
heavy neutrino discovery and low scale mass models. 
A comparison with future experiments is also included.

\subsection{Mass model scan}

We have randomly generated neutrino mass matrices and numerically diagonalised them.
The structure of the mass matrix is a generalised version of an ISS:
\begin{equation}
	\mathcal{M} = 
	\begin{pmatrix}
		0	& m_D^T	& 0	\\
		m_D	& \mu_R	& M^T_R	\\
		0	& M_R	& \mu_S
	\end{pmatrix}\ ,
\end{equation}
with two LNV submatrices, $\mu_R$ and $\mu_S$.
The number of physical parameters of a ISS\,$(a,b)$ mass matrix is $n_p = 7a + b +2\, a\,b$~\cite{Abada:2014vea}. %
We choose a basis in which $m_D$ has complex entries but three of which are real, $M_R$ is diagonal and real, and $\mu_S$ has a real diagonal %
without loss of generality.
If the matrix entries respect the hierarchy \mbox{$\mu_{R,S} \ll m_D \ll M_R$}, the mass spectrum in %
the LNC limit is principally given by the diagonal values of $M_R$.
We then perturb the matrix to achieve the three minimal ISS scenarios introduced above; %
the randomly generated mass matrix $\mathcal{M}$ %
is then diagonalised using the Jacobi Singular Value Decomposition (SVD) as implemented in the Eigen library~\cite{Eigen:2010}.
The Takagi decomposition, %
\begin{equation*}
	\hat{U}^T \mathcal{M}\, \hat{U} = \text{diag}(m_1, m_2, m_3, ...)\ ,
\end{equation*}
is retrieved starting from the SVD decomposition $\mathcal{M} = V \Sigma U^\dagger$, %
from which the singular values $\Sigma$ are the non-negative square roots of the eigenvalues of $\mathcal{M}^\dagger \mathcal{M}$ %
and the unitary matrix is $\hat{U} = U \rho^\dagger$, where $\rho = (U^T V)^\frac{1}{2}$ is a unitary phase matrix.

Only matrices satisfying the current constraints on heavy neutral fermions are taken in account.
The first requirement is that the eigenvalues must give the correct mass squared splittings %
compatible within 3$\sigma$ with the measured values~\cite{nufit}.
The condition of matching also the measured mixing angles is relaxed because %
the entries of the PMNS matrix, $\mathcal{U}$, are the result of the random structure of $m_D$ and $\mu_S$.
Constraints on the unitarity of the mixing matrix are applied instead.
The deviation from unitarity are quantified by the following Hermitian matrix:
\begin{equation}
	\varepsilon_{\alpha \beta} \equiv |\delta_{\alpha \beta} - (\mathcal{U}\,\mathcal{U}^\dagger)_{\alpha \beta}| = %
	\abs{\sum_{i=4}^n \hat{U}_{\alpha i} \hat{U}_{\beta i}^*}\ .
\end{equation}
The non-unitarity of the PMNS matrix has been assessed in various experiments, and the constraints depend upon the mass scale of averaged out neutrinos.
For neutrino masses below the GeV scale, but heavy enough to decouple from flavour oscillations, %
non-unitarity effects are tested in neutrino oscillation experiment as an overall normalisation.
If the neutrino mass is above the GeV scale, electroweak precision experiments provide strong constraints on non-unitarity.
The constraints are summarised below (from~\refrefs{Antusch:2008tz, Fernandez-Martinez:2016lgt, Blennow:2016jkn})
\begin{align*}
	&\varepsilon_{\alpha \beta} <
	\begin{pmatrix}
		\np{2.4e-2}	& \np{1.3e-2}	& \np{3.5e-2}	\\
		\cdot		& \np{2.2e-2}	& \np{6.0e-3}	\\
		\cdot		& \cdot		& \np{1.0e-1}
	\end{pmatrix}\quad 
	\text{if 10\,eV} \lesssim m_N \lesssim \text{1\,GeV}\ ,\\
	&\varepsilon_{\alpha \beta} <
	\begin{pmatrix}
		\np{1.3e-3}	& \np{1.2e-5}	& \np{1.4e-3}	\\
		\cdot		& \np{2.2e-4}	& \np{6.0e-4}	\\
		\cdot		& \cdot		& \np{2.8e-3}
	\end{pmatrix}\quad
	\text{if}\ m_N \gtrsim \text{1\,GeV}\ .
\end{align*}

The $\mu_R$ and $\mu_S$ entries of the ISS matrices naturally lead to lepton flavour and lepton number violating processes.
The most constrained process is the decay rate of \mbox{$\mu^+ \ra e^+\gamma$}, the branching ratio of which is given in \refeq{eq:meg}.
The current upper limit on the branching ratio is \np{4.2e-13}, but a future upgrade of the experiment %
foresees to reach a limit lower than \np{5e-14}.

Heavy neutrinos in a ISS model also contribute to the neutrinoless double beta decay.
The effective neutrino mass $m_{\beta\beta}$ receives further corrections with respect %
to the standard expression as
\begin{equation}
	m_{\beta\beta} \simeq \abs{\sum_i \hat{U}_{e i}^2 \frac{p^2\,m_i}{p^2 - m_i^2}}
\end{equation}
where $p^2 \simeq -\np{0.015}$\,GeV\tapi{2} is the typical virtual momentum of the exchanged neutrino.
The~contribution from masses above the 0.1\,GeV scale drops as $\flatfrac{1}{m^2_i}$ while it is constant for masses below~\cite{Blennow:2010th}.
It is interesting to note that the contributions given by pseudo-Dirac pairs are subject to partial cancellation, regulated by the LNV parameters.
In the LNC limit, the cancellation is maximum and the paired states do not take part in the $0\nu\beta\beta$ process.
The latest result from the KamLAND-Zen experiment~\cite{KamLAND-Zen:2016pfg} is interpreted as~\mbox{$m_{\beta\beta} < 61$\,meV}.

We find, for the first two ISS scenarios, that the allowed ranges span in the space %
\mbox{$m_D \sim 10^{[3,6]}$\,eV}, \mbox{$M_R \sim 10^{[6,15]}$\,eV}, $\mu_S$ and $\mu_R \sim 10^{[-4,1]}$\,eV.
We check that each matrix generated respects the \emph{naturalness condition} in the 't Hooft sense~\cite{tHooft:1980xss} %
and that the mass spectrum presents a mass state accessible by the DUNE experiment.
For the third ISS case, large entries of the sub-matrix $\mu_S$ are necessary to give the Majorana state a mass that %
can be probed by the experiment.
We find the ranges of \mbox{$m_D \sim 10^{[3,10]}$\,eV}, $M_R \sim 10^{[7,15]}$\,eV, $\mu_S \sim 10^{[4,9]}$\,eV to respect %
the constraints.
The hierarchy and naturalness conditions are relaxed in this case.
It is found that the block $\mu_R$ does not influence the final mass spectrum; %
it usually gives contribution to the light neutrino masses at the loop level, in a region below the GeV scale that has been already excluded by experiments.
The resulting points in the space $(m_N, |U_{\alpha N}|^2)$ are clustered together and the regions defined are overlaid in \reffig{fig:sensAll}.
Any combination of mass and mixing element inside these areas can be justified by a valid neutrino mass matrix %
which can explain the light neutrino masses and survive the experimental constraints.
The pseudo-Dirac pairs from the ISS\,(2,2) and ISS\,(2,3) scenarios give very similar regions, %
but Majorana states from the ISS\,(2,3) realisation can only be generated with very small couplings.
A type I seesaw band, corresponding to light neutrino mass between 20~meV and 200~meV, %
is plotted as well for comparison.

\subsection{Overall sensitivity}

\begin{figure}
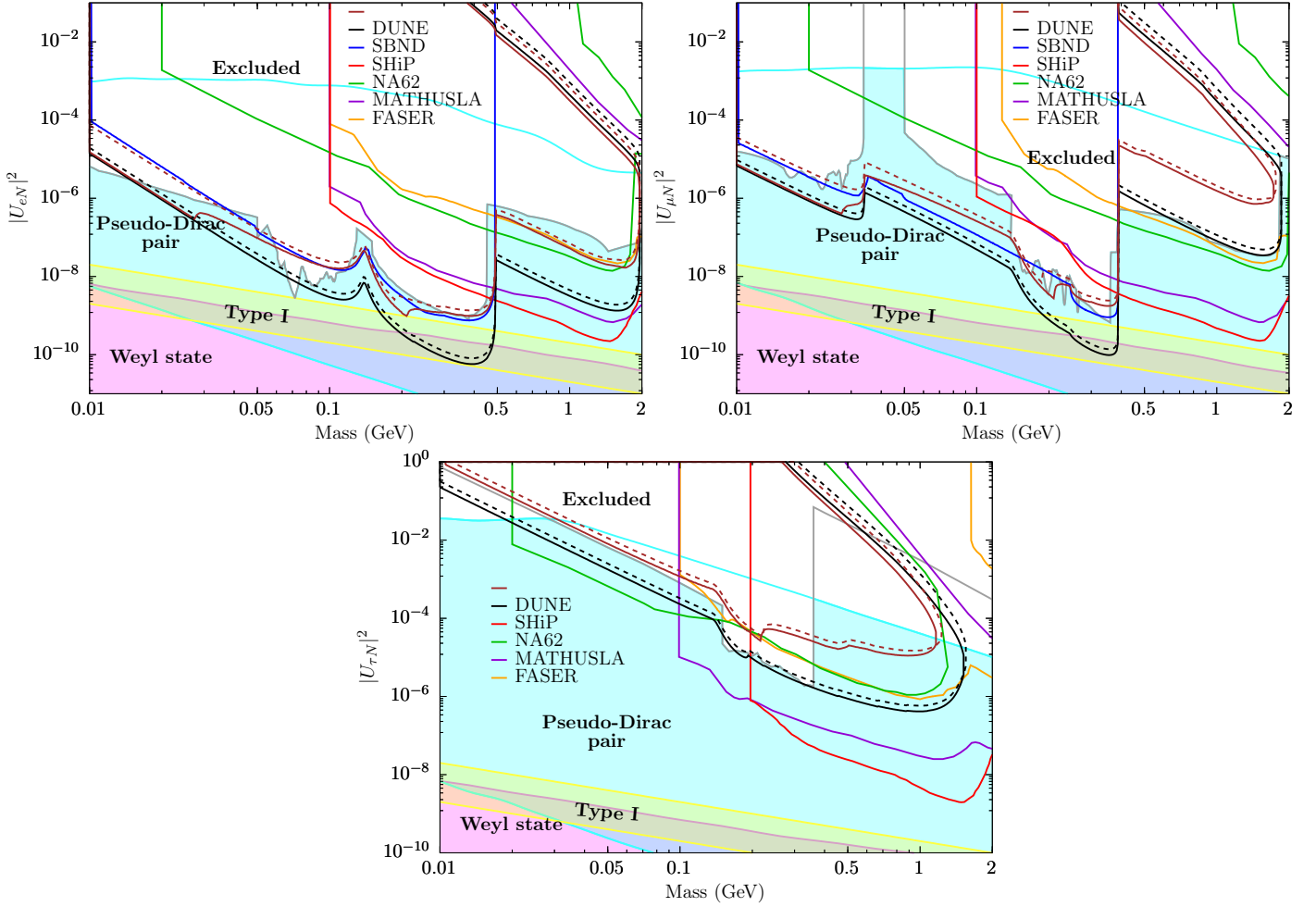

	\centering
	\noindent\makebox[\textwidth][c]{%
		\begin{minipage}{1.2\linewidth}
			\hspace{-1.5em}
			{\resizebox{0.5\linewidth}{!}{\input{DUNE_HNL_E_real.tex}}}
			{\resizebox{0.5\linewidth}{!}{\input{DUNE_HNL_M_real.tex}}}
		\end{minipage}
	}

	\vspace{0.5em}
	{\resizebox{0.6\linewidth}{!}{\input{DUNE_HNL_T_real.tex}}}
	\vspace{-0.2em}
	\footnotesize
	\caption{The 90\,\% C.L. sensitivity regions for dominant mixings %
		$|U_{e N}|^2$ (top left), $|U_{\mu N}|^2$ (top right), and $|U_{\tau N}|^2$ (bottom) are presented %
		combining results for channels with good discovery prospects (see text).
		The study is performed for Majorana neutrinos (solid) and Dirac neutrinos (dashed), %
		in the case of no background (black) and after the background analysis (brown).
		The region excluded by experimental constraints (grey) is obtained by combining the results from
		PS191~\cite{Bernardi:1985ny, Bernardi:1987ek}, %
		peak searches~\cite{Artamonov:2014urb, Britton:1992pg, Britton:1992xv, Aguilar-Arevalo:2017vlf, Aguilar-Arevalo:2019owf}, %
		CHARM~\cite{Vilain:1994vg}, NuTeV~\cite{Vaitaitis:1999wq}, DELPHI~\cite{Abreu:1996pa}, and T2K~\cite{Abe:2019kgx}, %
		with the lines reinterpreted for Majorana neutrinos (see \refref{Ruchayskiy:2011aa}).
		The sensitivity for DUNE ND (black) is compared to the predictions of future experiments, %
		SBN~\cite{Ballett:2016opr} (blue), %
		SHiP~\cite{Alekhin:2015byh} (red), NA62~\cite{Drewes:2018irr} (green), MATHUSLA~\cite{Curtin:2018mvb} (purple), %
		and FASER~\cite{Kling:2018wct} with 1\,m radius (orange).
		The shaded areas corresponds to possible neutrino mass models considered in this article: %
		the simulations of the ISS\,(2,2) and ISS\,(2,3) models where the lightest %
		pseudo-Dirac pair is the neutrino decaying in the ND (cyan); %
		the ISS\,(2,3) scenario when the single Majorana state is responsible for a signal (magenta); %
		the type~I seesaw scenario with a neutrino mass starting from \np{20}\,meV to \np{0.2}\,eV (yellow).}
	\label{fig:sensAll}
\end{figure}

We define the overall sensitivity of DUNE ND to the discovery of HNL as the combination of the sensitivities %
to some selected channels, presented in \reffig{fig:sensAll}.
These channels are $N \to \nu e^+ e^-$, $\nu e^\pm \mu^\mp$, $\nu \mu^+\mu^-$, $\nu\pi^0$, $e^\mp\pi^\pm$, and $\mu^\mp\pi^\pm$, %
and are preferred because of their good discovery prospect, for which backgrounds have also been studied.
They all give strong sensitivities, especially for masses below 0.5\,GeV, as shown in \refsec{sec:results}.
Their reach is due to high branching ratios and the HNL flux being more intense at such masses.
Also, the final state particles are all well-studied particles, most of which leave tracks in the detector %
that are easy to reconstruct, therefore allowing the background to be controlled with sufficient precision.
The neutrino spectrum component coming from the $D_s$ meson allows for weaker sensitivity %
to masses above the neutral kaon mass.
We conducted the sensitivity study for both scenarios, in which either a Majorana or a Dirac neutrino is the decaying~particle.

To appreciate the ND performance, we make a comparison with results of previous experiments, %
in particular PS191~\cite{Bernardi:1985ny, Bernardi:1987ek}, peak searches~\cite{Artamonov:2014urb, Britton:1992pg, Britton:1992xv}, %
CHARM~\cite{Vilain:1994vg}, NuTeV~\cite{Vaitaitis:1999wq}, DELPHI~\cite{Abreu:1996pa}, and T2K~\cite{Abe:2019kgx}.
We find that the DUNE ND can increase the bound on the electronic and muonic mixing elements %
for masses $m_N < m_{K^0}$ with respect to past experiments.
The constraint on the tauonic mixing is at least comparable with previous measurements.
For masses above, for which neutrino production relies on charm meson decays, the existing bounds %
are improved for the electronic mixing and the tauonic mixing, while a conservative result %
can be achieved in the muonic case.
We also overlay the prospects for the SBN programme~\cite{Ballett:2016opr}, %
NA62~\cite{Drewes:2018irr}, and the proposed SHiP~\cite{Alekhin:2015byh}, MATHUSLA~\cite{Curtin:2018mvb}, %
and FASER~\cite{Kling:2018wct} with 1\,m radius.
DUNE ND will give the best sensitivity for masses below the 0.5\,GeV in all channels, but the tauonic one.
However, anywhere the $D_s$ meson production is involved, the experiment cannot outperform the predicted %
sensitivity of the SHiP experiment which will deploy a 400\,GeV proton beam on a titanium-zinc-molybdenum alloy %
target, enhancing the production of charm and bottom mesons.
MATHUSLA will have a similar sensitivity, collecting particles from the High Luminosity LHC phase.
NA62 gives better results for the $|U_{\mu N}|^2$ mixing, but DUNE has a better sensitivity %
to the electron and tau channels.
FASER is comparable to NA62 in sensitivity, but it can reach regions of the parameter space beyond the 2\,GeV limit %
to which DUNE is not sensitive.
Comparing to previous similar studies, the sensitivities estimated in this analysis give stronger %
or at least comparable bounds than the ones in \refref{Krasnov:2019kdc}, 
where a different ND configuration is assumed, and no background study was performed.
More specifically, the limits on $|U_{eN}|^2$ are stronger, even considering the background events.
This is true also for the limits on $|U_{\mu N}|^2$, but only for masses below 500 MeV: %
in \refref{Krasnov:2019kdc} the sensitivity to masses above this threshold is enhanced by the contribution %
from B meson, which is not estimated in this study.
For the same reason, the limits on $|U_{\tau N}|^2$ prove to be comparable to our result, %
despite accounting only for the $D_s$ meson component.

We then compare the overall sensitivity to regions allowed by neutrino mass models.
In the electronic and muonic channels, DUNE ND will be sensitive to a large part of the pseudo-Dirac regions, %
corresponding to ISS\,(2,2) and ISS\,(2,3) models, %
part of which have been already excluded by past experiments.
DUNE will close the gap and put to test type I seesaw parameters, especially for HNL masses between 0.2 and 0.5\,GeV, %
starting to reach the region of ISS\,(2,3) with large lepton number violation.
For the tauonic channel, the experiment will probe only a small portion %
of pseudo-Dirac pairs from ISS\,(2,2) and ISS\,(2,3) models.
The sensitivity is not high enough to reach type I and Majorana state regions, which not even the dedicated experiment SHiP can.

\begin{figure}
	\centering
	{\resizebox{\linewidth}{!}{\input{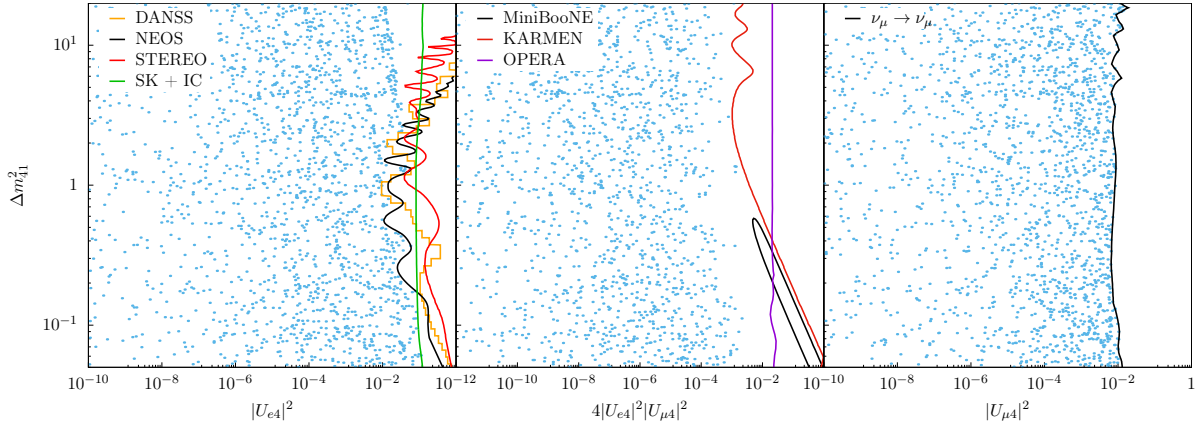}}}
	%
	\footnotesize
	\caption{One of the two ISS\,(2,3) realisations considered presents a Majorana state at masses comparable with SBL experiments.
		We show the results of the ISS\,(2,3) simulation (blue dots) for $\Delta m_{4 1}^2$ against combination of mixing angles and %
		the experimental result at 90\,\% C.L.: %
		$|U_{e 4}|^2$~(left) compared to DANSS~\cite{Alekseev:2018efk}, NEOS~\cite{Ko:2016owz}, STEREO~\cite{AlmazanMolina:2019qul}, %
		and Super-Kamiokande and IceCube combined~\cite{Dentler:2018sju};
		\mbox{$\sin^2 2\theta{_\mu e} = 4|U_{e 4}|^2|U_{\mu 4}|^2$} (middle) compared %
		to KARMEN2, OPERA and MiniBooNE~\cite{Aguilar-Arevalo:2018gpe},
		and $|U_{\mu 4}|^2$ (right) compared to a combined $\nu_\mu$ disappearance analysis~\cite{Dentler:2018sju}.
		Only the points generated by matrices which pass the experimental constraints are shown here.}
	\label{fig:sblosc}
\end{figure}

The ISS\,(2,3) scenario in which the pseudo-Dirac pair is accessible by the experiment also predicts a light %
Majorana state, the mass of which is controlled by the small LNV perturbations.
This entails the presence of a third mass splitting $\Delta m^2_{4 1}$, which could give %
an active-sterile oscillation signature in short baseline experiments.
In \reffig{fig:sblosc}, the new mass splitting is plotted against the mixings $|U_{e 4}|^2$, %
$|U_{\mu 4}|^2$ and the combination usually referred to as %
$\sin^2 2\theta_{\mu e} \equiv 4 |U_{e 4}|^2|U_{\mu 4}|^2$.
The mass splittings generated in the matrix scan span from %
$\Delta m^2_{3 1} \simeq \np{0.0025}$\,eV\tapi{2} up to \np{e4}\,eV\tapi{2}, and the squared mixings cover a large region, %
down to \np{e-14} for all the flavours.
The reactor anomalies could be soon excluded at the 90\,\% C.L. by the DANSS experiment~\cite{Alekseev:2018efk} %
and the allowed regions from LSND~\cite{Aguilar:2001ty} and %
MiniBooNE~\cite{Aguilar-Arevalo:2012fmn, Aguilar-Arevalo:2013pmq, Aguilar-Arevalo:2018gpe} %
require values of $\sin^2 2\theta_{\mu e} \gtrsim \np{e-3}$.
Given the results of the matrix scan, it is unlikely that one of the ISS\,(2,3) realisations %
considered in this work could link an heavy neutrino--like signal in DUNE ND and explain a short baseline anomaly at the same time, %
\enlargethispage{\baselineskip}
unless for sparse and very fine-tuned points.

\section{Conclusions}
\label{sec:conclusions}

Adding an arbitrary number of heavy neutral fermions is the simplest extension of the Standard Model which allows to address the neutrino mass origin.
These models are accompanied with a diverse and rich phenomenology, %
which can be tested by the next-generation neutrino experiments.
This is the case of low-scale seesaw mechanisms, such as the inverse seesaw %
which, depending on the realisation, allows Majorana or pseudo-Dirac heavy neutrinos with experimentally accessible masses.
In this paper, we have thoroughly investigated the phenomenological consequences of Majorana and Dirac states %
in light of searches of neutrino decays in beam dump experiments.
Production and decay modes have been computed using the helicity-spinor formalism, %
and all the formulae for differential decay rates and production scale factors are provided, for the first time, decomposed by helicity states.
We find agreement with previous studies, and hopefully settle down the dispute on different results.

We have shown that Dirac and Majorana neutrinos have different total decay width in NC processes %
and, in principle, measuring the rate could be a way of determining the nature of the initial state.
We put a lot of stress on the role of the helicity in these type of signatures: %
interesting differences appear between Majorana and Dirac neutrinos, which could be also %
exploited to determine the nature of the heavy singlet fermion.
The effect of the heavy neutrino helicity appears in the differential decay rate leading to different %
distributions of final state particles. %
For example, if the HNL are Majorana, two-body decays present an isotropic distribution for both helicity states, %
or, if Dirac, the angular distribution has a dependency proportional to $A\pm B \cos\theta$, %
with the sign depending on the helicity state.
We have also developed an effective evaluation of the heavy neutrino flux which, differently from a light neutrino flux, %
is not polarised to a single helicity state.
The production modes of a nearly-sterile neutrinos are sensitive to its helicity state, %
due to mass effects which can lead to enhancement of certain channels with respect to light neutrinos.
The two components of the neutrino flux behave therefore differently thanks to the dependency of decay distribution on the helicity.

We have studied the prospects for production and detection of HNL at the ND of the DUNE experiment.
The ND will be exposed to an intense neutrino beam and its exceptional reconstruction capabilities make it %
an ideal candidate for searches of heavy neutrino decays.
If at least one extra neutral state exists with a mass from few MeV to the GeV, %
the new singlet would be produced in the beam from mixing-suppressed meson and lepton decays.
It can subsequently decay inside the ND to the channels listed in \reftab{tab:decays}.
Thanks to the high energy of the beam, we have considered the possibility of testing neutrino masses heavier than the kaon mass.
We have carried out a simulation of $D_s$ meson production and decay, extending the analysis up to neutrino masses of 2\,GeV. %
More importantly, this has also allowed us to put constraints on $|U_{\tau N}|^2$ mixing, which is weakly bounded.

A background study was performed on decay channels with good detection prospects, %
defined by high branching ratios and clean detector signatures.
Due to the ND vicinity to the beam target, it is fundamental to suppress the overwhelming %
number of SM neutrino--nucleon interactions, which constitute the background for the rare signal of HNL decays.
Reconstruction of hadronic activity at the vertex and the multiplicity of final state particles are %
most of the time enough to distinguish between signal and background, reducing the latter down to~$\lesssim5$\,\%.
To further reduce unwanted events, simple kinematic cuts are applied thanks to the very forward distribution of %
decay in-flight events, additionally suppressing the background events to less than \np{5e-5} of the original number.
The rejection prescriptions are tuned to maintain an acceptable signal efficiency, which is between $\sim$40\,\% down to $\sim$20\,\%.
For all the other channels, no background study was performed, mainly because the final state particles %
are vector mesons which present experimentally challenging and specific signatures, the study of which %
was out of the scope of this work.
Combining the scaled flux components with the decay probabilities and signal efficiencies, %
we estimate the 90\,\% C.L.\ sensitivity of DUNE ND to all accessible channels, %
for both single and two dominant mixings.
For masses between 0.3 and 0.5\,GeV, the ND can probe mixing elements below \np{e-9} in most cases, %
reaching \np{e-10}, especially with two-body semi-leptonic channels for both $|U_{e N}|^2$ and $|U_{\mu N}|^2$.
Thanks to the $D_s$ meson production, neutrino masses above 0.5\, and up to 2\,GeV become accessible, %
as well as production and decay modes purely sensitive to the tau mixing.
In this case, the sensitivity does not exceed \np{e-8} for the electronic and muonic channels and %
\np{5e-6} for the tauonic channel. We point out that a large fraction of these parameters fall in the region relevant for the production of the baryon asymmetry via the ASR leptogenesis mechanism. 

Finally, we performed a random matrix scan of different ISS realisations to define regions of parameter space %
allowed by the model under consideration.
We identify three possible minimal cases that can provide good HNL candidates and at the same time address the lightness of the neutrino masses.
The first two correspond to an ISS\,(2,2) and an ISS\,(2,3) scenarios in which the heavy neutrino is part of the lightest pseudo-Dirac pair.
The third case is when strong LNV perturbations in a ISS\,(2,3) realisation give the Weyl state a mass accessible by the experiment.
We make sure that the matrices generated are in agreement with oscillation data on neutrino masses and satisfy the constraint %
imposed by other experiments on unitarity and lepton number violation.
We stress that DUNE will mostly---but not exclusively---be sensitive to pseudo-Dirac states.
In the region with strongest sensitivity, which is for masses just below 0.5\,GeV for $|U_{e N}|^2$ and %
below 0.4\,GeV for $|U_{\mu N}|^2$, the ND starts intersecting regions of the parameter space %
valid for a type I seesaw realisation or Majorana states in the ISS\,(2,3) scenario.
This might have consequences for the signal and analysis strategies adopted by the collaboration, %
according to the different topology of distribution between Majorana and pseudo-Dirac neutrinos.
\enlargethispage{\baselineskip}
In case of a discovery, some considerations can be drawn upon the nature of the new fermionic states.


\acknowledgments

The authors would like to thank Laura Fields and the DUNE collaboration for providing information on the fluxes.
PB and TB thank C\'edric Weiland for helpful comments on the ISS.
TB thanks Michele Lucente for useful tips on numerical scanning techniques and Richard Ruiz for discussions on open charm production.

This work has been supported by the European Research Council under ERC Grant ``NuMass'' (FP7-IDEAS-ERC ERC-CG 617143) %
and the European Union's Horizon 2020 research and innovation program under the Marie Sk\l{}odowska-Curie grant agreements %
No.~690575 (RISE InvisiblesPlus) and No.~674896 (ITN Elusives).
SP acknowledges partial support from the Wolfson Foundation and the Royal Society.

\appendix

\section{List of integrals and identities}
\label{app:integrals}

In presenting the differential and total decay rates in~\refsec{sec:decay} and~\ref{sec:production}, %
we have used a series of simplifying integrals and functions of the particle masses.
We report them jointly here.
The letters $x$, $y$, and $z$ denote squared ratios of masses, while $s$, $t$, and $u$ are the corresponding Mandelstam variables for three body decays.

\subsection{Decay widths}

In~\cite{Atre:2009rg}, the following functions are used to express the total rates of two-body decays  
\begin{align*}
	I_1 (x, y) &= \kallen(1, x, y)\, \qty[(1-x)^2 - y(1+x)]\ ,\\
	I_2 (x, y) &= \kallen(1, x, y)\, \qty[(1+x-y)(1+x+2y)-4x]\ ,
\end{align*}
and the rate of three-body decays can be expressed in terms of two more functions \cite{Atre:2009rg}, 
\begin{align*}
	I_1(x,y,z) &= 12 \int\limits_{\qty(\sqrt{x}+\sqrt{y})^2}^{\qty(1-\sqrt{z})^2}\!\! %
	\frac{\dd{s}}{s} (s-x-y)\,(1+z-s)\kallen(1,x,y)\kallen(1,s,z)\ ,\\
	I_2(x,y,z)&=24\,\sqrt{yz} \int\limits_{\qty(\sqrt{y}+\sqrt{z})^2}^{\qty(1-\sqrt{x})^2}\!\! %
	\frac{\dd{s}}{s}(1+x-s)\kallen(s,y,z)\kallen(1,s,x)\ .
\end{align*}
In this work we have introduced two differential generalisations of the two-body formulae,
\begin{align*}
	I^\pm_1 (x, y; \theta) &= \frac{1}{4\pi}\kallen(1, x, y) \qty[(1-x)^2 - y\,(1+x) \pm (x-1)\kallen(1, x, y) \cos\theta]\ ,\\
	I^\pm_2 (x, y;\theta) &= \frac{1}{4\pi}\kallen(1, x, y) \qty[(1+x-y)\,(1+x+2y)-4x \pm (x+2y-1)\kallen(1, x, y) \cos\theta]\ .
\end{align*}
Our expressions satisfy the normalisation conditions,  
\begin{align*}     
	\int_0^{2\pi} \!\! \dd{\varphi}\!\int_{-1}^1\!\!\dd{\cos\theta}\, I^\pm_1(x,y;\theta)&= I_1(x,y)\ ,\\
	\int_0^{2\pi} \!\! \dd{\varphi}\!\int_{-1}^1\!\!\dd{\cos\theta}\, I^\pm_2(x,y;\theta) &= I_2(x,y)\ . 
\end{align*}
We also note the following integrals which are necessary in deriving the total decay rate for the three-body leptonic modes, 
\begin{align}   
	\int\!\! \dd{s_1}\!\! \int\!\! \dd{s_2}\, (s_2-\xi^2_3)(1+\xi^2_4-s_2) &= \frac{I_1(0,\xi^2_3,\xi_4^2)}{12}\ ,\label{eq:threebody_int1}\\ 
	\int\!\! \dd{s_1}\!\! \int\!\! \dd{s_2}\, (s_1-\xi^2_4)(1+\xi^2_3-s_1) &= \frac{I_1(0,\xi^2_4,\xi_3^2)}{12}\ ,\label{eq:threebody_int2}\\ 
	\int\!\! \dd{s_1}\!\! \int\!\! \dd{s_2}\ 2\xi_3\,\xi_4(s_1+s_2-\xi^2_3-\xi^2_3) &= %
	\frac{I_2(0,\xi^2_3,\xi_4^2)}{12}\ ,\label{eq:threebody_int3}
\end{align}
where $\xi_i$ have the same meanings of \refeqs{eq:threebody_1}{eq:threebody_2}.

\subsection{Scaling factors for three body-decays}


Three-body lepton decays can produce neutrinos in two ways, depending on whether the neutrino mixes with %
the initial or with the final flavour.
The expressions presented in~\refsec{sec:production} make use of the following integrals:
\begin{align*}
	I^\pm_\ell(x, y, z) = 12 \int\limits_{\qty(\sqrt{x}+\sqrt{y})^2}^{\qty(1-\sqrt{z})^2}\!\!  \frac{\dd{s}}{s} %
	&\qty(1 + z - s) \qty[s - x - y \mp \kallen(s, x, y) ] \\
	&\times \kallen(s, x, y) \kallen(1, s, z)\ ,
\end{align*}
\begin{align*}
	I^\pm_{\cj{\ell}}(x, y, z) = 12 \int\limits_{\qty(\sqrt{x}+\sqrt{y})^2}^{\qty(1-\sqrt{z})^2}\!\!  \frac{\dd{s}}{s} %
	&\qty[1 + z - s \mp \kallen(1, s, z) ] \qty(s - x - y) \\
	&\times \kallen(s, y, z) \kallen(1, s, z)\ .
\end{align*}
When averaging over the helicity states, these two functions become identical and, because of symmetry crossing, %
also identical to the integral $I_1(x, y, z)$, expressed above.


In~\refsec{sec:production}, the three-body decay rate of pseudoscalar meson requires the following integral:
\begin{align*}
	I^\pm_h(x, y, z) = \int\limits_{\qty(\sqrt{x}+\sqrt{y})^2}^{\qty(1-\sqrt{z})^2}\!\!
	\dd{s} \int_{t_-}^{t_+}\!\! %
	\dd{t} \qty[ F^2 A^\pm(s, t) + G^2 B^\pm(s, t) - \Re(F^* G)\,C^\pm(s, t) ]\ , \\
	\text{with}\quad t_\pm = x + z + \frac{ (1-s-z)(s - y + x) \pm \kallen(s, y, z)\kallen(1, s, z) }{2 s}\ ,
\end{align*}
where $F$ and $G$ are convenient combinations of hadronic form factors $f^{(h,h')}$.
From lattice QCD considerations, form factors should carry the correct Clebsch-Gordan, %
but here we drop them as they are irrelevant when studying scale factors.
The combinations $F$ and $G$ are
\begin{align*}
	F &= 2\ f_+^{(h,h')}(u) = f_+^{(h,h')}(0)%
	\qty(1 + \lambda^{(h,h')}_+ \frac{u}{x})\ , \\
	G &= f_+^{(h,h')}(u) - f_-^{(h,h')}(u) = f_+^{(h,h')}(0)%
	\qty[1 + \lambda^{(h,h')}_+ \frac{u}{x} - %
	\qty(\lambda^{(h,h')}_+ - \lambda^{(h,h')}_0) \qty( 1 + \frac{1}{x})]\ ,
\end{align*}
with $\lambda$ parametrising the linear dependence~\cite{PDG} of the form factors %
with respect the momentum transfer between the two mesons, $u$, %
directly connected to the other Mandelstam variables, $s$ and $t$:
\begin{equation*}
	u = 1 + x + y + z - s - t\ .
\end{equation*}
The values of $\lambda_{+,0}$ is determined experimentally~\cite{PDG}.
The functions $A$, $B$, and $C$ are
\begin{align*}
	A^\pm(s, t) &= \frac{1}{2}(1 + y - t) \qty[ 1 + z - s \mp \kallen(1, z, s) ] - %
	\frac{1}{2}\qty[u - y - z \mp \kallen(u, y, z)]\ , \\
	B^\pm(s, t) &= \frac{1}{2}(y + z) (u - y - z) + 2 y z \mp (y - z) \frac{\kallen(u, y, z)}{2}\ , \\
	C^\pm(s, t) &= z (1 + y - t) + \qty[y \pm \frac{\kallen(u, y, z)}{2}] (1 + z - s)\ . 
\end{align*}
When summing over helicity states, the kinematic simplifies to
\begin{align*}
	A(s, t) &= (1 + y - t)( 1 + z - s) - (u - y - z)\ , \\
	B(s, t) &= (y + z) (u - y - z) + 4\, y\, z\ , \\
	C(s, t) &= 2\, z\, (1 + y - t) + 2\, y\, (1 + z - s)\ .
\end{align*}

\section{Polarised $N\to\nu \ell_\alpha^-\ell^+_\beta$ distributions}
\label{app:threebody_dist}

\subsection{Dirac $\nu_i$} 
The coefficients for a Dirac neutrino decay are given by 
\begin{align*}
	C^\nu_1 &= C^\nu_4= \sum_{\gamma =e}^\tau |U_{\gamma i}|^2 \qty[\delta_{\alpha\beta} g_L^2+ %
	\delta_{\gamma\alpha}(1+\delta_{\alpha\beta}g_L)]\ ,\\ 
	C^\nu_2 &= C^\nu_5= \delta_{\alpha\beta}\,g^2_R\sum_{\gamma = e}^\tau |U_{\gamma i}|^2\ ,\\ 
	C^\nu_3 &= C^\nu_6= \delta_{\alpha\beta}\,g_R\sum_{\gamma = e}^\tau |U_{\gamma i}|^2(\delta_{\gamma\beta} + g_L)\ ,
\end{align*}
where the chiral couplings for charged leptons are given by $g_L = -\frac{1}{2} + \sin^2\theta_\text{W}$ 
and \mbox{$g_R = \sin^2\theta_\text{W}$}.

\subsection{Dirac $\overline{\nu}_i$}  
The coefficients for the Dirac antineutrino decay---which involve some vital minus signs compared %
to the neutrino case---are given by 
\begin{align*}
	C^{\cj{\nu}}_1 &= -C^{\overline{\nu}}_4 = \delta_{\alpha\beta}\,g^2_R\sum_{\gamma = e}^\tau |U_{\gamma i}|^2\ ,\\
	C^{\cj{\nu}}_2 &= -C^{\overline{\nu}}_5 = \sum_{\gamma =e}^\tau |U_{\gamma i}|^2 %
	\qty[ \delta_{\alpha\beta} g_L^2+ \delta_{\gamma\beta}(1+\delta_{\alpha\beta}g_L)]\ ,\\
	C^{\cj{\nu}}_3 &= -C^{\overline{\nu}}_6 = \delta_{\alpha \beta}\, g_R\sum_{\gamma = e}^\tau |U_{\gamma i}|^2(\delta_{\alpha\gamma} + g_L)\ ,
\end{align*}
where the chiral couplings $g_L$ and $g_R$ have the same meaning.

\subsection{Majorana $N_i$}  
The amplitude for Majorana decay is the sum of the Dirac neutrino and Dirac antineutrino amplitudes given above:%
\footnote{In general, there are interference terms between ``neutrino'' and ``antineutrino'' diagrams; %
	however all such contributions are suppressed by the mass scale of the outgoing light neutrino, %
	which is taken to be zero in these calculations.}
\begin{equation*}
	|A_\pm|^2 = |A^\nu_\pm|^2 + |A^{\cj{\nu}}_\pm|^2\ .
\end{equation*} 
%
%
Crucially, this means that the coefficients in the isotropic terms are the sum of those for a neutrino and antineutrino while the coefficients in the angular terms are the difference, leading to cancellations.
All in all, we find
\begin{equation*}
	|A_\pm|^2 = |A_0|^2 \pm |A_1|^2\ , 
\end{equation*}
with the coefficients
\begin{align*}
	C_1&=C_1^\nu + C_1^{\cj{\nu}} = \sum_{\gamma =e}^\tau |U_{\gamma i}|^2 \qty[ (g_L^2+g_R^2)\delta_{\alpha\beta} + %
	\delta_{\gamma\alpha}(1+\delta_{\alpha\beta}g_L)]\ ,\\
	C_2&=C_2^\nu + C_2^{\cj{\nu}} = \sum_{\gamma =e}^\tau |U_{\gamma i}|^2 \qty[ (g_L^2+g_R^2)\delta_{\alpha\beta} + %
	\delta_{\gamma\beta}(1+\delta_{\alpha\beta}g_L)]\ ,\\
	C_3&=C_3^\nu + C_3^{\cj{\nu}} = 2\delta_{\alpha\beta}\,g_R\sum_{\gamma = e}^\tau |U_{\gamma i}|^2(\delta_{\alpha\gamma} + g_L)\ ,\\
	C_4&=C_1^\nu -C_1^{\cj{\nu}} = \sum_{\gamma =e}^\tau |U_{\gamma i}|^2 \qty[ \delta_{\alpha\beta}(g_L^2-g_R^2) + %
	\delta_{\gamma\alpha}(1+\delta_{\alpha\beta}g_L)]\ ,\\
	C_5&=C_2^\nu - C_2^{\cj{\nu}} = -\sum_{\gamma =e}^\tau |U_{\gamma i}|^2 \qty[ \delta_{\alpha\beta}(g_L^2-g_R^2) + %
	\delta_{\gamma\beta}(1+\delta_{\alpha\beta}g_L)]\ ,\\
	C_6&= C_3^\nu - C_3^{\cj{\nu}} = 0\ .
\end{align*}
Note that for the three-body decays, the decay is not isotropic in the Majorana limit; however, the quantity $g_L^2 - g_R^2 \approx 0.02$, suppresses the angular terms in the pure NC case.

\section{Open charm production}
\label{sec:opencc}

\begin{figure}
	\centering
	\begin{fmffile}{qq}
		\begin{fmfgraph*}(80,50)
			\fmfset{arrow_len}{2.5mm}
			\fmfleft{q1,q0}
			\fmfright{c1,c0}
			\fmflabel{$q$}{q0}
			\fmflabel{$\cj{q}$}{q1}
			\fmflabel{$c$}{c0}
			\fmflabel{$\cj{c}$}{c1}
			\fmf{fermion}{q0,vl,q1}
			\fmf{gluon}{vl,vr}
			\fmf{fermion}{c1,vr,c0}
		\end{fmfgraph*}
	\end{fmffile}
	\hspace{0.2em}
	\raisebox{2em}{,}
	\hspace{0.2em}
	\begin{fmffile}{gg_0}
		\begin{fmfgraph*}(80,50)
			\fmfset{arrow_len}{2.5mm}
			\fmfleft{g1,g0}
			\fmfright{c1,c0}
			\fmflabel{$g$}{g0}
			\fmflabel{$g$}{g1}
			\fmflabel{$c$}{c0}
			\fmflabel{$\cj{c}$}{c1}
			\fmf{gluon}{g0,vl}
			\fmf{gluon}{g1,vl}
			\fmf{gluon}{vr,vl}
			\fmf{fermion}{c1,vr,c0}
		\end{fmfgraph*}
	\end{fmffile}
	\hspace{0.1em}
	\raisebox{2em}{$+$}
	\hspace{0.1em}
	\begin{fmffile}{gg_1}
		\begin{fmfgraph*}(80,50)
			\fmfset{arrow_len}{2.5mm}
			\fmfleftn{g}{2}
			\fmfrightn{c}{2}
			\fmf{fermion}{c1,vt,vb,c2}
			\fmf{gluon}{g1,vt}
			\fmf{gluon}{g2,vb}
			\fmflabel{$g$}{g1}
			\fmflabel{$g$}{g2}
			\fmflabel{$\cj{c}$}{c1}
			\fmflabel{$c$}{c2}
		\end{fmfgraph*}
	\end{fmffile}
	\hspace{0.1em}
	\raisebox{2em}{$+$}
	\hspace{0.1em}
	\begin{fmffile}{gg_2}
		\begin{fmfgraph*}(80,50)
			\fmfset{arrow_len}{2.5mm}
			\fmfleftn{g}{2}
			\fmfrightn{c}{2}
			\fmf{gluon}{g1,vt}
			\fmf{gluon}{g2,vb}
			\fmf{phantom}{c2,vb,vt,c1}
			\fmf{fermion,tension=0}{c1,vb,vt,c2}
			\fmflabel{$g$}{g1}
			\fmflabel{$g$}{g2}
			\fmflabel{$\cj{c}$}{c1}
			\fmflabel{$c$}{c2}
		\end{fmfgraph*}
	\end{fmffile}
	\vspace{1em}
	\caption{These are the four diagrams contributing to the hard process in open charm production.
		The diagrams with gluons in the initial state interfere with each other giving rise to %
		cross terms in the colour structure.}
	\label{fig:parton}
\end{figure}
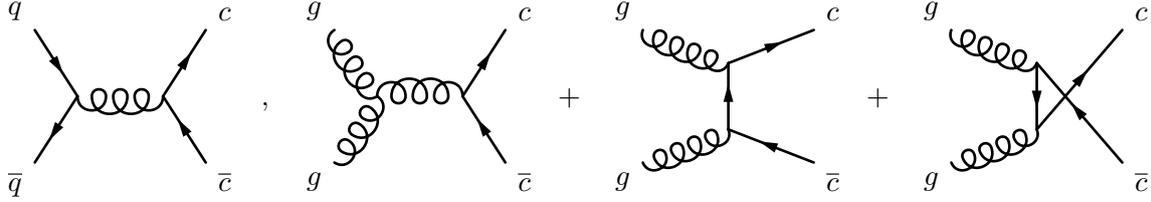

Following the same procedure as the one described in~\refref{Alekhin:2015byh}, %
we estimate the number of strange $D$ mesons to be
\begin{equation}
	\mathcal{N}_{D_s} = \frac{\sigma_{c \cj{c}}}{\sigma_{p A}} f_{D_s} = (2.8 \pm 0.2) \times \np{e-6}\ ,
\end{equation}
where $\sigma_{c \cj{c}} = \np{12}\pm\np{1}$\,\textmu b is the proton-target open charm cross section, %
$\sigma_{p A} = \np{331.4}\pm\np{3.4}$\,mb is the total inelastic proton-target on carbon ($A =$ \tapi{12}C)~\cite{RamanaMurthy:1975vfu} %
cross section, and $f_{D_s} = \np{7.7}\,\%$ is the $D_s$ fragmentation fraction~\cite{Abramowicz:2013eja}.
We calculate the open charm production cross section at the leading order in perturbation theory, with a %
graphite fixed target and a 80\,GeV proton $p$.
The correct process to consider is the proton--nucleon interaction, therefore %
\begin{equation*}
	\sigma_{c \cj{c}} \equiv \sigma(pA \ra c \cj{c} + X) \approx A\,\sigma(pN \ra c \cj{c} +X)\ , %
\end{equation*}
using the correct Parton Distribution Function (PDF) for a bound nucleon $N$ in the nucleus $A$.
There are four diagrams, shown in \reffig{fig:parton}, that contributes to the cross section, %
but three of them interfere with each other.
These cross sections are well-known SM calculations and can be found in~\refref{PDG}.
The integrated cross section is:
\begin{multline}
	\sigma(pN \ra c \cj{c} + X) = \int_{\tau_0}^1 \dd{x_1} \int_{\frac{\tau_0}{x_1}}^1 \dd{x_2} \int \dd{\Omega} %
	\bigg[\qty(f_{g/p}^1\, f_{g/A}^2 + f_{g/p}^2\, f_{g/A}^1) \dv{\sigma_{gg \ra c \cj{c}}}{\Omega} \\
	+\sum_{q=u,d,s} \qty(f_{q/p}^1\, f_{\cj{q}/A}^2 + f_{q/p}^2\, f_{\cj{q}/A}^1 + %
	f_{\cj{q}/p}^1\, f_{q/A}^2 + f_{\cj{q}/p}^2\, f_{q/A}^1) \dv{\sigma_{q\cj{q} \ra c \cj{c}}}{\Omega}\bigg]\ ,
\end{multline}
with $\tau_0 = \hat{s}_0 / s$ and $\hat{s}_0$ being the threshold energy at the partonic level and %
$s = 2 m_p ( m_p + E_p)$ is the centre of mass energy, given that $m_p \simeq m_n$.
The partonic structure of the nucleus is described by the functions $f_{\rho/\eta}^i = f_{\rho/\eta}(x_i, M_F)$, %
which are interpreted as the probability of finding a parton $\rho$ in the particle $\eta$ %
carrying a $x_i$ fraction of the momentum of $\eta$, at the energy scale $M_F$.	
The two momentum fractions are related by $x_1\,x_2\,s = \hat{s}$, where the hat symbol denotes the energy %
of the parton-level process.

We adopt a factorisation scale of $M_F = 2.1\, m_c$ for the computation of $\sigma_{c \cj{c}}$, 
while the renormalisation scale of $\alpha_s$ is set to $\mu_R = 1.6\, m_c$, and the charm mass has the value %
\mbox{$m_c = (\np{1.28}\pm\np{0.03})$\,GeV}.
The integration is regulated for $|\cos \theta| < 0.8$, with $\theta$ the angle in the centre of mass frame.
The theoretical curve in Fig.~7.4(a) of~\refref{Alekhin:2015byh} was used to check our evaluation, %
and it was successfully reproduced up to NLO corrections.
For~the calculation we employed LHAPDF~\cite{Buckley:2014ana} and the nCTEQ15 PDF set~\cite{Kovarik:2015cma}, %
resulting in $\sigma_{pA \ra c \cj{c}} = (\np{12}\pm1)\,\mu$b, for an 80\,GeV protons on a graphite target.

\section{Background reduction}
\label{sec:appbackground}

We performed a background study only for the decay channels with an important discovery potential, %
and these are $N \to \nu e^+ e^-$, $\nu e^\pm \mu^\mp$, $\nu \mu^+\mu^-$, $\nu\pi^0$, $e^\mp\pi^\pm$, and $\mu^\mp\pi^\pm$. 
In~order to reject background events, conservative event selection cuts are outlined using the differences %
between kinematic properties of the final state particles from neutrino--nucleon interactions and %
from the rare HNL in-flight decays.
Simulations of signal events with a given mass inside either the LArTPC or the MPD are input to a %
channel-specific algorithm that discards low energy events and %
defines limits on angular and transverse momentum distributions.
The algorithm aims at keeping an integrated signal efficiency $\widehat{W}_d$ greater than 30\,\%, where 
\begin{equation*}
    \widehat{W}_d = \int \dd{E} W_d(E)\ ,
\end{equation*}
where the signal efficiency $W_d(E)$ is introduced in \refsec{sec:numevt}.

As an example of the selection process, we present here the results of the analysis for %
a heavy neutrino with mass $m_N = 450$\,MeV.
In the following tables the number of background events is reported %
in the form ``$\mathcal{X} \to \mathcal{Y}\ \mathcal{Z}$'', %
where $\mathcal{X}$ is the per mille (\np{e-3}) fraction of background events %
from mis-identification and $\mathcal{Y}$ and $\mathcal{Z}$ are fractions of irreducible background %
after the application of selection cuts to respectively Majorana and Dirac neutrino simulations.
When the value $\np{0.000}$ is shown, we mean that less than one background event per million is expected.
The average $\langle\nu\rangle$ is computed by weighting the flux components contribution to the background, %
using the respective interaction rates as weights, reported in Tab.~\ref{tab:rate}.
To obtain the number of background events, each fraction must be multiplied by the number of %
SM neutrino--nucleon interactions expected in the ND during the experiment lifetime.
We assume that the $\nu_\tau$ and $\cj{\nu}_\tau$ components are not responsible for background events, %
therefore only the $\nu_e$, $\nu_\mu$, and $\cj{\nu}_\mu$ components are studied.
The last row of the tables show the signal efficiency of the selection cuts.

We group the studied channels in three categories, which have similar kinematic features: %
two-body decay, which are semi-leptonic, three-body decay channels, which are purely leptonic instead, and %
decays which can be only detected via photon reconstruction.

\subsection{Two-body decays}

The two-body decays $N \to e^\pm \pi^\mp$ and $N \to \mu^\pm \pi^\mp$ are the most promising channels for the detection %
of a heavy neutrino, being the decay mode with the highest branching ratios.
Since all final state particles are charged, direct information on the parent particle in easily reconstructed, %
as for instance the mass of the decaying neutrino, which is the invariant mass of the process
\begin{equation*}
    m_N^2 = s = m_\ell^2 + m_\pi^2 + 2E_\ell E_\pi - 2|\vb{p}_\ell| |\vb{p}_\pi| \cos \theta\ ,
\end{equation*}
where $\theta$ is the opening angle between the lepton and the pion.
In a two-body decay, the two particles are emitted back-to-back in the neutrino reference frame, %
so in the laboratory frame the relative position on the perpendicular plane is mostly preserved %
and $(\phi_\ell - \phi_\pi)$ is expected to be close to $\pm \pi$.
Despite these distinctive signatures, these two channels are the ones with most background events, %
coming from charged-current interactions of $\nu_e$, $\nu_\mu$, and $\cj{\nu_\mu}$ in which %
additional pions can be easily emitted in coherent or deep inelastic scatterings.
Background events typically peak at low energies and present more isotropic angular distributions.
Therefore, a tight energy threshold on the energies of the charge particles is imposed to accept %
70\,\% of the signal events and a threshold on the energy of the reconstructed neutrino is defined %
by 90\,\% of the retained events.
A cut is also placed on the reconstructed $m_N$ to retain 80\,\% of signal events, %
as well as an upper limit on the transverse momenta and angles to the beamline %
and a lower and an upper limit on the separation angle between the charged particles.
After the cuts are applied, the background events are reduced up to a factor of 2500, %
and the signal efficiency are $\sim$35\,\% for the electronic channel and $\sim$40\,\% %
for the muonic channel, with little difference (respectively 1\,\% and 3\,\%) %
between Dirac or Majorana selection windows.

\begin{center}
\smallskip
	\small
	\begin{tabular}{cr@{~}c@{~~}cr@{~}c@{~~}c}
	\toprule

	& \multicolumn{3}{c}{$N\to e^\mp \pi^\pm$}		& \multicolumn{3}{c}{$N\to \mu^\mp \pi^\pm$}	\\

	\cmidrule(lr){2-4} \cmidrule(lr){5-7}   

	& & Majorana		& Dirac	 & & Majorana	& Dirac	\\

	\cmidrule(lr){2-4} \cmidrule(lr){5-7} 

	$\nu_e$         &\np{19.090}~~$\to$ & \np{0.015} & \np{0.015}	&\np{ 0.007}~~$\to$ & \np{0.000} & \np{0.000}	\\
	$\nu_\mu$       &\np{ 0.027}~~$\to$ & \np{0.000} & \np{0.000}	&\np{25.030}~~$\to$ & \np{0.011} & \np{0.012}	\\
	$\cj{\nu}_\mu$  &\np{ 0.025}~~$\to$ & \np{0.000} & \np{0.000}	&\np{29.822}~~$\to$ & \np{0.046} & \np{0.053}	\\

	\cmidrule(lr){2-4} \cmidrule(lr){5-7}

	$\langle\nu\rangle$		&\np{ 0.239}~~$\to$ & \np{0.000} & \np{0.000}	&\np{24.302}~~$\to$ & \np{0.013} & \np{0.014}	\\

	\cmidrule(lr){2-4} \cmidrule(lr){5-7}

	$\widehat{W}_{\ell\pi}$&		& 36.4\,\%	& 35.2\,\%	&		& 43.3\,\%	& 40.2\,\% \\

	\bottomrule
	\end{tabular}
\medskip
\end{center}

\subsection{Three-body decays}

The three-body decays studied are $N\ra \nu e^- e^+$, $N\ra \nu e^\mp \mu^\pm$, and $N\ra \nu \mu^- \mu^+$.
The event selection in this case is more challenging compared to two-body decays event, due the loss of the light neutrino %
which precludes the reconstruction of the decaying HNL, and so cuts as rigorous cannot be defined.
However, since two charged leptons are needed to identify these channels, the resulting background rate, %
from mis-identified photons (from $\pi^0$ decays) and long-track pions, is low.
Even in this case, only high energy events are considered, but with a lower threshold on the charged lepton energies.
The invariant mass of the two leptons has as upper limit $m_N$ and this constrain helps with reducing the background.
Lower and upper limits are also defined for the transverse momenta, as well as separation angles from the beamline.

The background events are reduced from a factor of 40 up to a factor of 200, with the selection requirements %
for Dirac neutrinos being more effective.
The signal efficiency results to be better (6--8\,\% better) for Majorana neutrinos in the $N\ra \nu e^- e^+$ and $\mu^-\mu^+$ %
channels, whereas the Dirac neutrino have give a better efficiency in the $N\ra e^\mp \mu^\pm$ channel.
High efficiency and low background make these three channel competitive for HNL discovery, despite having %
lower branching ratio and so weaker sensitivity.

\begin{center}
\smallskip
	\noindent\makebox[\textwidth][c]{%
\begin{minipage}{1.02\textwidth}
	\small
	\begin{tabular}{cr@{~}c@{~~}cr@{~}c@{~~}cr@{~}c@{~~}c}
	\toprule

 & \multicolumn{3}{c}{$N\to \nu e^- e^+$}	& \multicolumn{3}{c}{$N\to \nu e^\mp \mu^\pm$}	& \multicolumn{3}{c}{$N\to \nu \mu^- \mu^+$} \\

	\cmidrule(lr){2-4} \cmidrule(lr){5-7}  \cmidrule(lr){8-10} 

	& & Majorana		& Dirac	 & & Majorana	& Dirac & & Majorana & Dirac	\\

	\cmidrule(lr){2-4} \cmidrule(lr){5-7}    \cmidrule(lr){8-10}

	$\nu_e$         &\np{0.190}~~$\to$ & \np{0.003} & \np{0.002}  &\np{0.078}~~$\to$ & \np{0.002} & \np{0.002}  &\np{0.000}~~$\to$ & \np{0.000} & \np{0.000} \\
	$\nu_\mu$       &\np{0.193}~~$\to$ & \np{0.001} & \np{0.000}  &\np{0.092}~~$\to$ & \np{0.000} & \np{0.000}  &\np{0.081}~~$\to$ & \np{0.001} & \np{0.001} \\
	$\cj{\nu}_\mu$  &\np{0.224}~~$\to$ & \np{0.003} & \np{0.002}  &\np{0.160}~~$\to$ & \np{0.000} & \np{0.000}  &\np{0.090}~~$\to$ & \np{0.008} & \np{0.006} \\
                                                                                                                                                                  
	\cmidrule(lr){2-4} \cmidrule(lr){5-7}    \cmidrule(lr){8-10}
	$\langle\nu\rangle$		&\np{0.168}~~$\to$ & \np{0.001} & \np{0.000}  &\np{0.090}~~$\to$ & \np{0.000} & \np{0.000}  &\np{0.022}~~$\to$ & \np{0.000} & \np{0.000}\\

	\cmidrule(lr){2-4} \cmidrule(lr){5-7}    \cmidrule(lr){8-10}

	$\widehat{W}_{\nu\ell\ell}$	&	& 63.4\,\%	& 55.4\,\%	&	& 68.6\,\%	& 71.2\,\% &	& 74.0\,\%	& 68.4\,\%	\\

	\bottomrule
	\end{tabular}
\end{minipage}
}
\medskip
\end{center}

\subsection{EM--detected decays}

The semi-leptonic decay $N \ra \nu \pi^0$ may only be identified by a correct photon reconstruction, %
since the neutral pion decays almost 100\,\% of the time in two photons.
This particle is produced in NC1$\pi^0$ interactions and deep inelastic scattering interactions.
Background events occur if only two final state photons from the neutral pion decay %
are above detection threshold and properly reconstructed with an invariant mass equal to $m_{\pi^0}$.
The energy of the reconstructed pion is the best discriminant against background events, %
thanks to their high energy.
Lower and upper limits can be placed on the $\pi^0$ transverse momentum and angle with the beamline, %
but also a threshold on the energy of the photons as well as an upper limit on their angular distributions %
help define the kinematics of the event.
The residual background for this channel is the highest among the ones studied: only reduction factors up to 130 can be achieved, %
with a notable difference between selection cuts for Majorana and Dirac HNL decays, the latter ones being more strict.
The signal efficiency is $\sim$46\,\% for Majorana and $\sim$42\,\% for Dirac.
It is, however, one of the decay modes with the highest branching ratio, %
and with advanced and dedicated techniques~\cite{Ankowski:2008aa, Back:2012wc}
the background rejection can be improved.

\begin{center}
\smallskip
	\small
	\begin{tabular}{cr@{~}c@{~~}c}
	\toprule

	& \multicolumn{3}{c}{$N\to \nu \pi^0$}	\\

	\cmidrule(lr){1-4}

	& & Majorana		& Dirac	 \\

	\cmidrule(lr){2-4} 

	$\nu_e$         &\np{4.135}~~$\to$ & \np{0.058}	& \np{0.048}	\\
	$\nu_\mu$       &\np{5.862}~~$\to$ & \np{0.053}	& \np{0.039}	\\
	$\cj{\nu}_\mu$  &\np{7.428}~~$\to$ & \np{0.179}	& \np{0.138}	\\

	\cmidrule(lr){2-4} 

	$\langle\nu\rangle$		&\np{5.797}~~$\to$ & \np{0.061}	& \np{0.045}	\\

	\cmidrule(lr){2-4} 

	$\widehat{W}_{\nu\pi^0}$	& & 46.3\,\%	& 42.3\,\%	 \\

	\bottomrule
	\end{tabular}
\medskip
\end{center}


\end{document}